\documentclass[twocolumn]{aastex63}
\usepackage{amsmath}
\usepackage{enumerate}
\usepackage{enumitem}   
\usepackage{hyperref}
\usepackage{color}
\usepackage{subfigure}
\usepackage{color}
\usepackage{courier}
\usepackage{mathtools}
\usepackage{threeparttable}
\usepackage{enumitem}
\usepackage{multirow}
\usepackage{CJK}
\defcitealias{Lin17}{Paper I} 	

\received{}
\revised{}
\accepted{}
\submitjournal{}

\shortauthors{Pan et al.}

\begin{document}

\begin{CJK}{UTF8}{bsmi}
\title{SDSS-IV MaNGA:  The Nature of an Off-galaxy H$\alpha$ Blob -- A Multi-wavelength View of Offset Cooling in a Merging Galaxy Group}

\correspondingauthor{Hsi-An Pan}
\email{pan@mpia.de}

\author{Hsi-An Pan (潘璽安)}
\affiliation{Institute of Astronomy and Astrophysics, Academia Sinica, No. 1, Section 4, Roosevelt Road, Taipei 10617, Taiwan}
\affiliation{Max-Planck-Institut f\"ur Astronomie, K\"onigstuhl 17, D-69117 Heidelberg, Germany}

\author{Lihwai Lin}
\affiliation{Institute of Astronomy and Astrophysics, Academia Sinica, No. 1, Section 4, Roosevelt Road, Taipei 10617, Taiwan}

\author{Bau-Ching Hsieh}
\affiliation{Institute of Astronomy and Astrophysics, Academia Sinica, No. 1, Section 4, Roosevelt Road, Taipei 10617, Taiwan}

\author{Micha{\l}~J.~Micha{\l}owski}
\affiliation{Astronomical Observatory Institute, Faculty of Physics, Adam
Mickiewicz University, ul.~S{\l}oneczna 36, 60-286 Pozna{\'n}, Poland}

\author{Matthew S. Bothwell}
\affiliation{Cavendish Laboratory, University of Cambridge, 19 J. J. Thomson Avenue, Cambridge CB3 0HE, UK}
\affiliation{University of Cambridge, Kavli Institute for Cosmology, Cambridge, CB3 0HE, UK}

\author{Song Huang}
\affiliation{Department of Astronomy and Astrophysics, University of California Santa Cruz, 1156 High St., Santa Cruz, CA 95064, USA}
\affiliation{Kavli-IPMU, The University of Tokyo Institutes for Advanced Study, the University of Tokyo}
\affiliation{Department of Astrophysical Sciences, Peyton Hall, Princeton University, Princeton, NJ 08540, USA}

\author{Alexei V. Moiseev}
\affiliation{Special Astrophysical Observatory, Russian Academy of Sciences, Nizhnij Arkhyz, 369167, Russia}
\affiliation{Space Research Institute, Russian Academy of Sciences, Profsoyuznaya ul. 84/32, Moscow 117997, Russia}

\author{Dmitry Oparin}
\affiliation{Special Astrophysical Observatory, Russian Academy of Sciences, Nizhnij Arkhyz, 369167, Russia}

\author{Ewan O'Sullivan}
\affiliation{Harvard-Smithsonian Center for Astrophysics, 60 Garden Street, Cambridge, MA 02138, USA}

\author{Diana M. Worrall }
\affiliation{HH Wills Physics Laboratory University of Bristol, Tyndall Avenue,  Bristol BS8 1TL,  UK}

\author{Sebasti\'an  F. S\'anchez}
\affiliation{Instituto de Astronom\'ia, Universidad Nacional Aut\'onoma de M\'exico, A. P. 70-264, C.P. 04510, M\'exico, D.F., Mexico}

\author{Stephen Gwyn}
\affiliation{NRC-Herzberg Astronomy and Astrophysics, National Research Council of Canada, 5071 West Saanich Road, Victoria, British Columbia V9E 2E7, Canada}

\author{David R. Law}
\affiliation{Space Telescope Science Institute, 3700 San Martin Drive, Baltimore, MD 21218, USA}

\author{David V. Stark}
\affiliation{Physics and Astronomy, Haverford College, Haverford, PA}

\author{Dmitry Bizyaev}
\affiliation{Apache Point Observatory and New Mexico State University, Sunspot, NM, 88349, USA}
\affiliation{Sternberg Astronomical Institute, Moscow State University, Moscow, Russia}

\author{Cheng Li }
\affiliation{Tsinghua Center of Astrophysics \& Department of Physics, Tsinghua University, Beijing 100084, China}

\author{Chien-Hsiu Lee}
\affiliation{NSF's National Optical-Infrared Astronomy Research Laboratory, Tucson, AZ, USA}

\author{Hai Fu}
\affiliation{Department of Physics \& Astronomy, University of Iowa, Iowa City, IA 52242, USA}

\author{Francesco Belfiore}
\affiliation{European Southern Observatory, Karl-Schwarzchild-Str. 2, Garching bei München, D-85748, Germany}

\author{Kevin Bundy}
\affiliation{Department of Astronomy and Astrophysics, University of California, Santa Cruz, 1156 High Street, Santa Cruz, CA 95064, USA}

\author{Jos\'e G. Fern\'andez-Trincado}
\affiliation{Instituto de Astronom\'ia y Ciencias Planetarias, Universidad de Atacama, Copayapu 485, Copiap\'o, Chile}

\author{Joseph Gelfand}
\affiliation{NYU Abu Dhabi, United Arab Emirates}
\affiliation{NYU Center for Cosmology and Particle Physics, New York, NY 10003, USA}

\author{S\'ebastien Peirani}
\affiliation{Universit\'e C\^ote d'Azur, Observatoire de la C\^ote d'Azur, CNRS, Laboratoire Lagrange, Nice, France}
\affiliation{Institut d'Astrophysique de Paris, CNRS \& UPMC, UMR 7095, 98 bis Boulevard Arago, F-75014 Paris, France}

\begin{abstract} 
Galaxies in dense environments, such as groups and clusters,  experience various  processes by which galaxies  gain and  lose gas.
Using data from the SDSS-IV MaNGA survey, we previously reported the discovery of a giant (6 -- 8 kpc in diameter) H$\alpha$ blob, Totoro,  about 8 kpc away from a pair of galaxies  (Satsuki and Mei) residing in a galaxy group which is experiencing a group-group merger.
Here, we combine interferometric $^{12}$CO(1--0) molecular gas data, new wide-field H$\alpha$, $u$-band data, and published  X-ray data to determine  the origin of the blob.
Several scenarios are discussed to account for its multi-wavelength properties, including (1)  H$\alpha$ gas being stripped from galaxy Satsuki  by ram-pressure; (2) a separated low-surface-brightness galaxy; (3)  gas being ejected or ionized by an active galactic nucleus (AGN); and (4) a cooling intra-group medium (IGM).
Scenarios  (1) and (2) are less favored by  the present data.
Scenario (3) is also less likely as  there is no  evidence for an active ongoing  AGN in the host galaxy.
We find that the CO (cold) and H$\alpha$ (warm) gas coexist with X-ray  (hot) structures; moreover, the derived cooling time is  within the regime where molecular and  H$\alpha$  gas are expected. 
The coexistence of gas with different temperatures  also agrees with that of  cooling gas in other systems.  
Our multi-wavelength results strongly suggest that the CO and H$\alpha$ gas are the product of cooling from the IGM at its current location, i.e., cooling has occurred,  and may be ongoing, well outside the host-galaxy core.

\end{abstract} 


\section{Introduction}
\label{sec_intro}

The optical colors of galaxies are characterized by a bimodality, with early-type galaxies residing on the red sequence and late-type  galaxies  populating the blue cloud \citep[e.g.,][]{Str01,Bal04,Bra09}. 
The transition  from  blue  to  red galaxies is known to be driven by a   decrease in star formation \citep[e.g.,][]{Whi12,Tay15}.
Since galaxies require gas to fuel star formation \citep[e.g.,][]{Sai17},  a knowledge of the gain and loss of gas provides a huge leap in the understanding of galaxy evolution.

Many galaxies reside in gravitationally bound groups or clusters, where a high volume density of galaxies is found.
A natural consequence of clustering of galaxies  is frequent  interactions  between  galaxies and the environment  in which they live \citep[e.g.,][]{Fer14}.
Such interactions can stimulate the  gain and loss and heating and cooling of gas in various ways and  control the star formation processes in galaxies.
For example, interaction between galaxies can potentially trigger nuclear activity and nuclear starbursts that generate energetic gas outflow/winds to the environments \citep[e.g.,][]{Lar78,Sil98,Spr05,Ell08,Hop08,Fer14}.
Besides, when  a  galaxy  moves at a high speed through a high density region, a complex hydrodynamical interaction of  its  interstellar medium (ISM) and the surrounding hot intracluster/intergalactic  medium takes place.
A large fraction of  gas could be removed  from  a    galaxy   if the ram-pressure is strong enough to overcome the gravitational force \citep{Gun72}.
These gas removal processes can  lead to the interruption of  star formation activity.

On the other hand, fresh supplies of gas can sustain star formation in galaxies.
Galaxies can gain gas through accretion of small companions \citep[][]{Bon91,Lac93}. 
Moreover, cold gas can form from the cooling  of  the  hot  intracluster  medium  (ICM) or intragroup medium (IGM) and  be accreted to galaxies \citep{Ega06,Ode08}.
Such  process can rejuvenate  early-type and S0 galaxies by supplying gas into their centers.

\begin{figure*}
	\begin{center}
		\subfigure[]{\label{fig_Halpha}\includegraphics[scale=0.465]{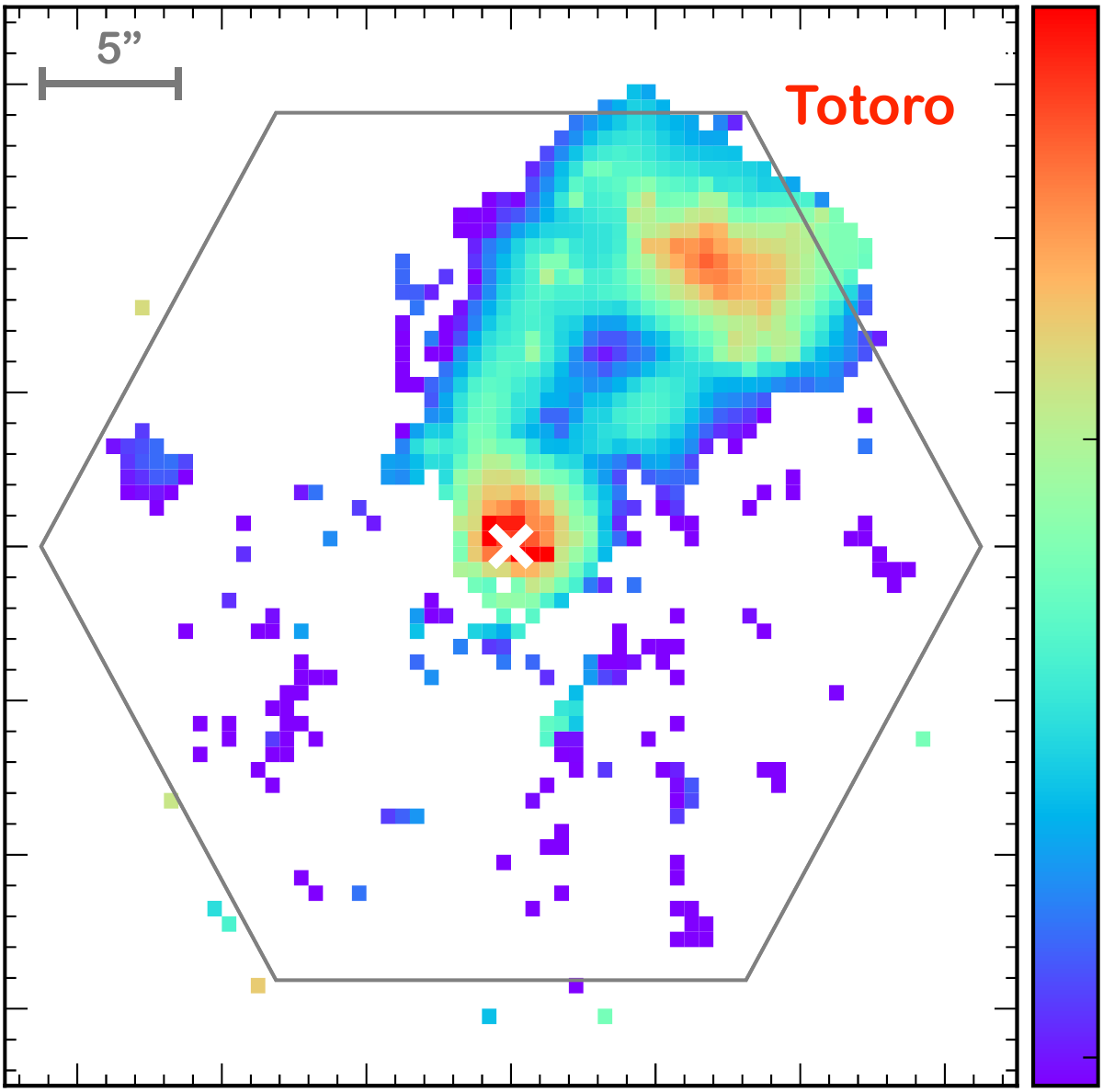}}
		\subfigure[]{\label{fig_cluster_zoom}\includegraphics[scale=0.4]{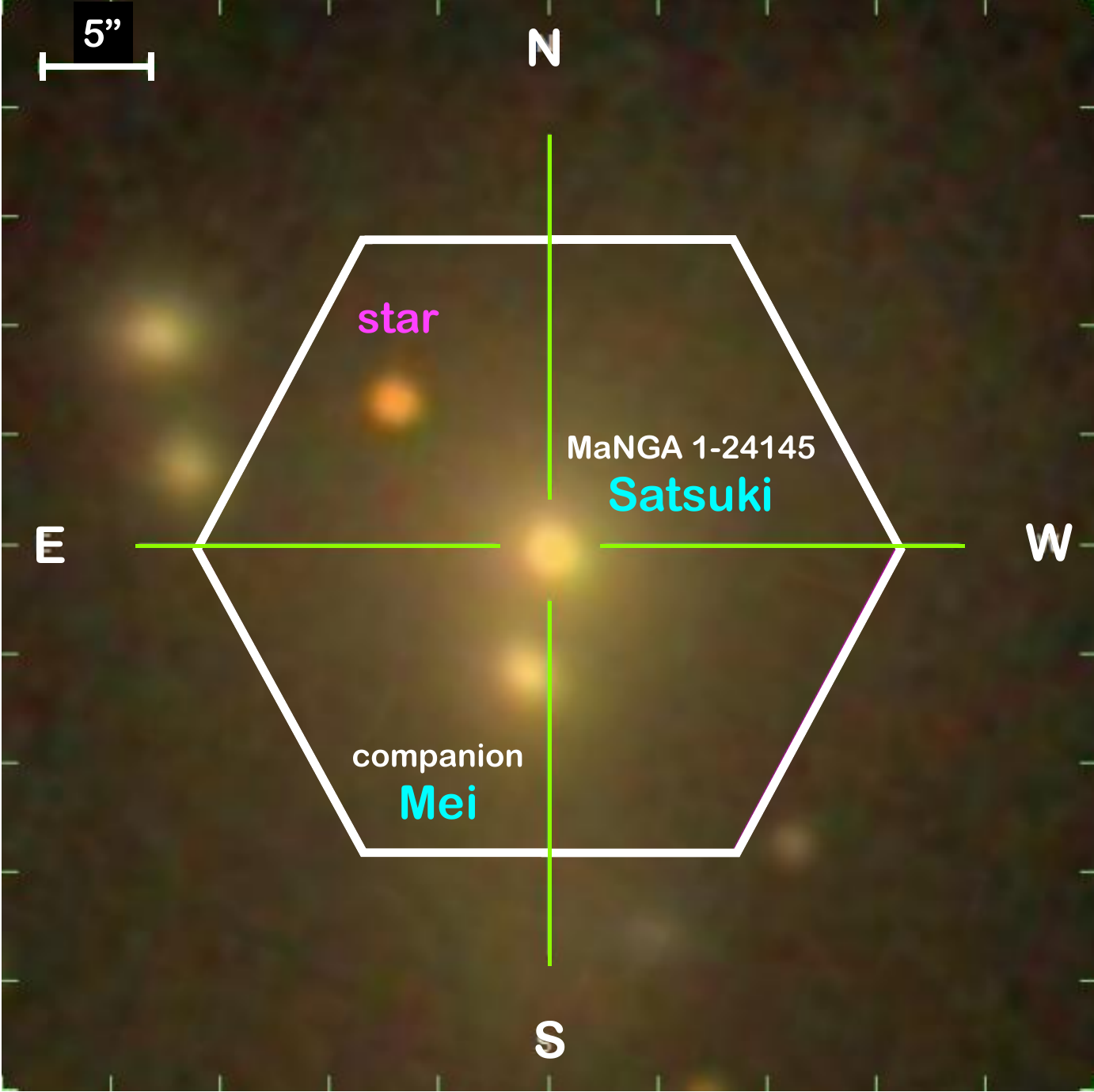}}
		\subfigure[]{\label{fig_cluster}\includegraphics[scale=0.4]{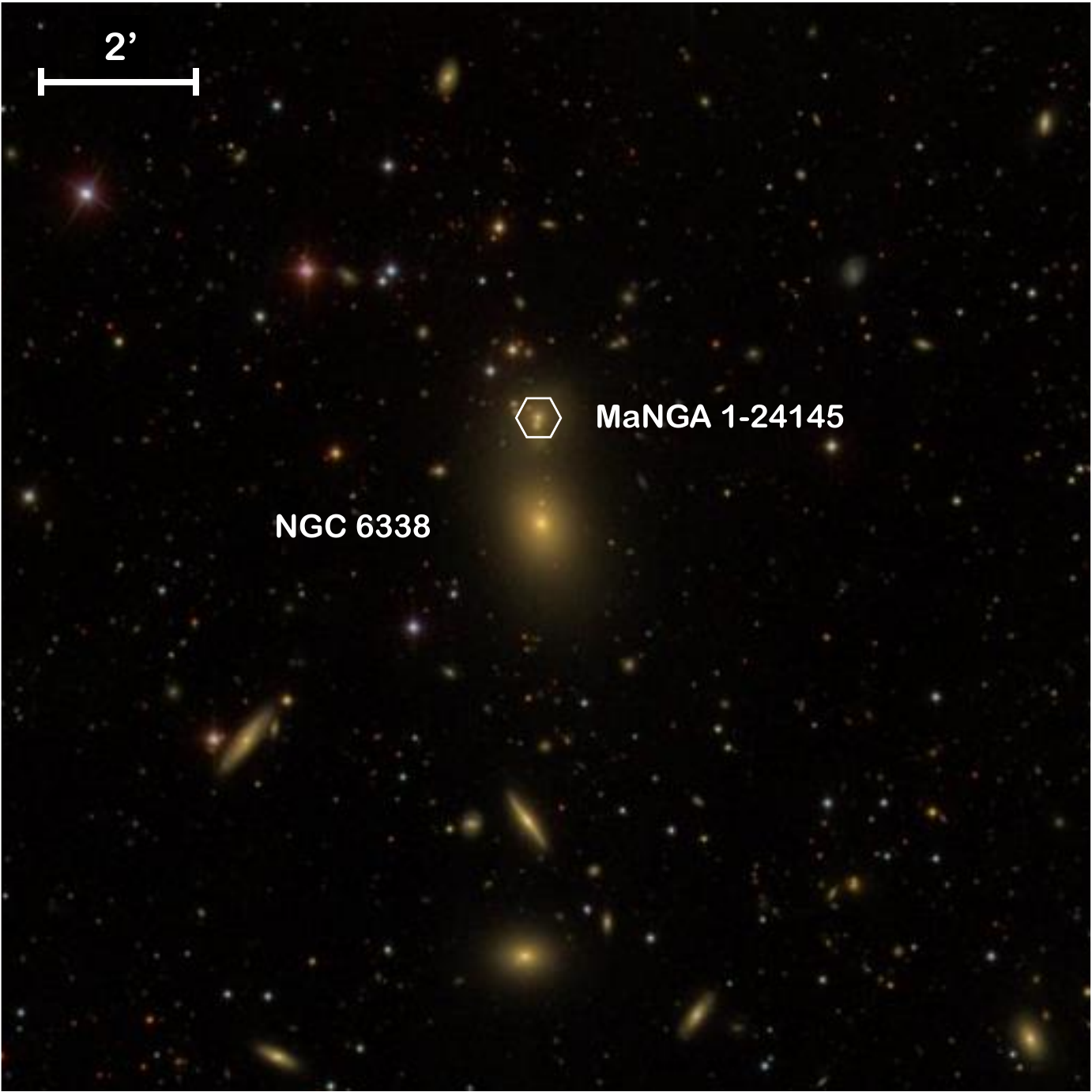}}

	\end{center}
	\caption{Optical images of MaNGA 1-24145 (nicknamed  ``Satsuki'') and the environment in which it lives. The hexagons show the coverage of  MaNGA  bundle field of view. Satsuki was observed with the 127 fiber bundle of MaNGA; the hexagon is $\sim$ 32.5$\arcsec$ in diameter. (a) MaNGA H$\alpha$ map of MaNGA 1-24145. An H$\alpha$ blob  (nicknamed ``Totoro'') is about 8 kpc northwest of MaNGA 1-24145. The unit of the map is 10$^{-16}$ erg s$^{-1}$ cm$^{-2}$. The data extend to regions just outside the hexagon because of the dithering.  (b) The SDSS $gri$ composite image centering on MaNGA 1-24145. The south companion  (nicknamed  ``Mei'') is also within the MaNGA field of view. 
	(c) The SDSS $gri$ composite image of MaNGA 1-24145 and the nearby galaxies.
	}
	\label{fig_clusters}
\end{figure*}

In our previous paper, \citet[][hereafter \citetalias{Lin17}]{Lin17}, we report the discovery of a  giant ionized gaseous (H$\alpha$) blob associated with a dry merger system (i.e., mergers between two early-type galaxies;   Figure   \ref{fig_Halpha} and \ref{fig_cluster_zoom}) residing in a galaxy group (Figure \ref{fig_cluster}).
 The ionized gaseous blob is offset from both of the  galactic nuclei. 
The  gaseous blob might be a result of galaxy interactions, active galactic nucleus (AGN) activity, or a galaxy interacting with or being accreted by  the dry merger. 
In any of these cases, we may be witnessing an ongoing gain and/or loss of gas in these galaxies.

The H$\alpha$ blob was identified from the first-year MaNGA survey \citep[Mapping Nearby Galaxies at APO;][]{Bun15}, part of SDSS-IV \citep{Bla17}.
The H$\alpha$ blob (nicknamed ``Totoro'') is $\sim$ 3 -- 4 kpc in radius and is $\sim$ 8 kpc away (in projection) from the host galaxy MaNGA target 1-24145  (or MCG+10-24-117; nicknamed  ``Satsuki''; 17$^\mathrm{h}$15$^\mathrm{h}$23.26$^\mathrm{s}$, +57$^{\circ}$25$\arcmin$58.36$\arcsec$, stellar mass $M_{\ast}$ $\approx$ 10$^{11}$ M$_{\sun}$).
There is no distinct optical continuum counterpart at the position of Totoro.
The mass of the ionized gas is 8.2 $\times$ 10$^{4}$ M$_{\sun}$ (\citetalias{Lin17}).
The SDSS image indicates that the host galaxy Satsuki has a companion galaxy (nicknamed
``Mei'') located to the south east of Satsuki (Figure \ref{fig_cluster_zoom}). 
The companion is also within the hexagon bundle field of view (FoV; $\sim$ 32.5$\arcsec$ in diameter) of MaNGA.
These two galaxies are at similar redshifts ($z$ $\approx$ 0.03), with their line of sight velocity differing by $\sim$ 200 km s$^{-1}$.
Both galaxies are elliptical and thus form a dry (gas-poor) merger, also known as VII Zw 700. 

On the large scale, the dry merger (Satsuki and Mei)  is located at the overlap region of a group-group merger  \citep{Osu19}.
Satsuki and Mei are associated with the less massive northern component, while 
the nearby large elliptical galaxy NGC 6338 is  the brightest galaxy of the  more massive group (Figure \ref{fig_cluster}).
The dry merger  (Satsuki and Mei) and NGC 6338  are separated by $\sim$ 42 kpc in projection and by $\sim$ 1400 km s$^{-1}$ in velocity.
The two merging groups are expected to form a galaxy cluster in the future.
The merger velocity of the two  groups is as large as 1700 -- 1800 km s$^{-1}$, making this one of the most violent mergers yet observed between galaxy groups \citep{Osu19}.

Our  H$\alpha$ blob Totoro has also been observed  by \cite{Osu19} using the APO 3.5-m telescope (see their Figure 10), but only the high-luminosity, main blob region was detected. 
\cite{Osu19}  also show that the H$\alpha$ gas of NGC 6338 consists of three diffuse filaments in the southeast and northwest quadrants, extending out to $\sim$ 9 kpc from the nucleus.
The H$\alpha$ filaments have also been revealed in  previous observations by $HST$ and the CALIFA\footnote{Calar Alto Legacy Integral Field Area Survey \citep{San12}.} survey \citep{Mar04,Pan12,Gom16}.
Moreover, \cite{Osu19} show that both NGC 6338 and VII Zw 700 contain potential X-ray cavities, which would indicate past AGN jet activity.
In addition, both systems are observed to contain cool IGM structures correlated with the H$\alpha$ emission \citep[see also][]{Pan12}.

We diagnosed the physical properties of Satsuki and Tororo using MaNGA data in  \citetalias{Lin17}:
\begin{itemize}
\item \emph{kinematics}: the ionized (H$\alpha$) gas component reveals that   there is a moderate velocity variation ($\leqslant$ 100 km s$^{-1}$)  from the position of  Satsuki along the connecting arms to Totoro, but there is no  velocity gradient across Totoro itself and the velocity and velocity dispersion across the blob are low ($\ll$ 100 km s$^{-1}$; Figure 7 of \citetalias{Lin17}).
\item \emph{excitation state}: the Baldwin-Phillips-Terlevich (BPT) emission line diagnostics \citep{Bal81}  indicate LI(N)ER-type excitations for Satsuki  and a composite (LI(N)ER-HII  mix\footnote{Shocks can also  lead to line ratios  occupying the composite regions. However, the analysis of shock and photoionization mixing models of \citetalias{Lin17} shows that the shocks are unlikely to be the dominant mechanism that is responsible for ionization of Totoro.}) for Totoro.
\item \emph{gas metallicity}:  metallicity\footnote{Most of the metallicity calibrators can only be applied for those regions  in which ionization is dominated by star  formation, and may not be applicable to regions with ionization parameters or ISM pressure different from typical HII regions. In  \citetalias{Lin17}, we adopted  the ``N2S2H$\alpha$''    calibrator, which is suggested to be less sensitive to the ionization parameters \citep{Dop16}. However, caution is still needed when interpreting the derived metallicity.}  of the gas around Satsuki is close to the solar value; on the other hand, the  metallicity  of Totoro is higher than that of Satsuki by 0.3 dex.

\end{itemize}

Several possibilities for  Totoro were raised in \citetalias{Lin17}, including (1)  the gas  being ram-pressure stripped from Satsuki, (2) a galaxy interacting with the dry merger  (Satsuki and Mei), and (3) gas being ejected or ionized  by an AGN associated with Satsuki.
However,  the data in \citetalias{Lin17}  are not sufficient to provide strong  constraints on the  nature of Totoro.

In this paper, we present new observations of this system and determine the most plausible origin of Totoro.
Our new observations include:
\begin{enumerate}[label=(\roman*)]
\item \emph{wide-field H$\alpha$ image}: to reveal the distribution of ionized gas at larger scale (i.e.,  beyond the FoV of the MaNGA bundle),  
\item \emph{$u$-band observation}: to constrain the ionizing source, i.e., star formation or not, and to search for the possible continuum counterpart of the H${\alpha}$ blob, and 
\item \emph{molecular gas in $^{12}$CO(1--0)}: to constrain the amount and distribution of cold, i.e., potentially star-forming, gas.
\end{enumerate}
Moreover, in addition to the  scenarios raised in \citetalias{Lin17}, \cite{Osu19}  argue for the ``cooling gas'' hypothesis in Satsuki, therefore, we also use
\begin{itemize}
	\item[(iv)] \emph{X-ray data} from \cite{Osu19} to constrain the properties of the surrounding hot medium.
\end{itemize}

The paper is organized as follows. 
Section \ref{sec_data} describes the observations and our data reduction.
The results are presented and discussed in Section \ref{sec_results}.
In Section \ref{sec_summary} we summarize our results and list our conclusions. 

Throughout this study, we assume a  cosmology with $\Omega_\mathrm{m}$ $=$ 0.3, $\Omega_\mathrm{\Lambda}$ $=$ 0.7, and $H_{0}$ $=$ 70 km s$^{-1}$ Mpc$^{-1}$.  We use a Salpeter stellar initial mass function.

\section{Data}
\label{sec_data}

\subsection{MaNGA}

MaNGA, the largest   integral field spectroscopy (IFS) survey of the nearby Universe  to date, observed  $\sim$ 10,000 galaxies with a median redshift ($z$) of 0.03.
The details of the MaNGA survey, the integral-field-unit (IFU)  fiber system,  the sample selection, observing strategy, and the data reduction and analysis pipelines are explained in  \citet{Dro15},  \citet{Wak17}, \citet{Law15},  \citet{Law16}, and \citet{Wes19}, respectively, and also summarized in \citetalias{Lin17}.


The MaNGA data used in this work were reduced using the MPL-7 version \citep[corresponding to SDSS data release 15,][]{Agu19}  of the MaNGA data reduction pipeline.
An earlier version of the pipeline (MPL-4) was used for \citetalias{Lin17}. 
The differences  between  the pipeline products of  MPL-4 and MPL-7 are negligible for the current work.
The spectral-line fitting is carried out using the Pipe3D pipeline \citep{San16a,San16b}.
Details of the fitting procedures are described in \cite{San16a,San16b} and summarized in \citetalias{Lin17}.
The method described in \cite{Vog13} is used to compute the reddening using the Balmer decrement at each spaxel.

\subsection{Wide-field  H$\alpha$ data}
Deep optical images were taken at  the prime focus of the 6-m telescope of the Special Astrophysical Observatory of the Russian Academy of Sciences (SAO RAS) with the SCORPIO-2  multimode focal reducers \citep{Afa11}. 
A narrow-band filter AC6775 (the central wavelength CWL $=$ 6769\AA, the bandwidth FWHM $=$ 15\AA) covers the spectral region around the redshifted H$\alpha$ emission line. Two middle-band filters FN655 (CWL $=$ 6559\AA,  FWHM $=$ 97\AA) and FN712 (CWL $=$ 7137\AA,  FWHM $=$ 209\AA) were used to obtain the blue and red  continuum images. We combined the data taken during two nights 05/06 and 06/07 Mar 2017 with  seeing of 1.3 -- 1.5$\arcsec$. The total exposure time depended on  filter FWHM: 7800, 1300, and 780 sec in the filters AC6775, FN655 and FN712, respectively.  
The detector,  CCD E2V 42-90 (2K $\times$ 4.5K), operated in the bin 2 $\times$ 2 read-out mode provides 0.35$\arcsec$/px scale in the 6.1$\arcmin$ field of view. 

The data reduction was performed in a standard way for  SCORPIO-2 direct image processing with IDL-based software \citep[see, for instance,][]{Sit15}. 
The underlying stellar continuum from the H$\alpha$ image was subtracted using the linear combination of the images in the filters FN655 and FN712. 
The astrometry grid was created using the Astrometry.net project web-interface\footnote{http://nova.astrometry.net/} \citep{Lan10}.

We emphasize that the purpose of the wide-field observation is  to reveal the distribution of  H$\alpha$ gas outside of the MaNGA FoV,  in particular, to look for a potential tidal tail(s) on the other side of Totoro.  
To be in line with the measurements and discussions in \citetalias{Lin17}, the MaNGA H$\alpha$ map will be used by default throughout this paper, and the wide-field H$\alpha$ map will be used only where specifically mentioned in the text.
The flux of the new narrow-band  image has been calibrated by the MaNGA spectral line data to avoid systematic errors related with narrow-band filter calibration  (e.g., only one standard star per night, mismatch between the redshifted lines and a peak of the filter transmission curve,  etc.). 
This forces the flux of the narrow-band  image to be consistent with that of MaNGA spectral line data.

\subsection{$u$-band Data}
The $u$-band observation was taken with the wide-field  imaging facility MegaCam   with a 1$^{\circ}$ field of view at the Canada–France–Hawaii Telescope (CFHT) from April 27th to June 24th in 2017 (PI: L. Lin; project ID:
17AT008).
The total exposure time is 12,000 seconds. 
The MegaCam data were processed and stacked via MegaPipe  \citep{Gwy08}. 
The final image has a limiting mag of 26.3 mag (1$\arcsec$ aperture in radius).

\subsection{Molecular Gas (CO) Data}

We mapped the $^{12}$CO($J$ $=$ 1 $\rightarrow$ 0; 115.2712 GHz) emission  with the NOrthern Extended Millimeter Array (NOEMA) at Plateau de Bure (PI: L. Lin; project ID: S16BE001).
The  full width at half power of the primary beam of each NOEMA antenna at the $^{12}$CO(1-0) frequency is $\sim$ 50$\arcsec$, sufficiently large to cover Satsuki, Mei, and Totoro.
The observations were spread across 11 nights  from  May 25th to July 18th of 2015  with 5 or 7 antennas in D configuration. 
The total on-science-source time is $\sim$ 22 hours.
The shortest possible baseline length is $\sim$ 24m, and therefore sources larger than about 15$\arcsec$ might be resolved out. 
Nonetheless, in Section \ref{sec_results}, we will see that the distribution of molecular gas matches the H$\alpha$ extremely well.


Data reduction, calibration, and imaging were performed with the CLIC and MAPPING software of GILDAS\footnote{http://www.iram.fr/IRAMFR/GILDAS} using standard procedures. 
Images were reconstructed using  natural weighting to preserve maximal sensitivity.  
The resultant synthesized  beam    is 3.4$\arcsec$ $\times$ 2.3$\arcsec$ (PA $=$ 73.3$^{\circ}$), with an effective beamsize of 2.8$\arcsec$.
 The aim of the observation is to constrain the amount of  molecular gas traced by $^{12}$CO(1-0), therefore a relatively low  spectral resolution (53.6 km s$^{-1}$) was proposed to maximize the detection probability. 
The rms noise in the $^{12}$CO(1-0) cube is 0.31 mJy beam$^{-1}$   per 53.6 km s$^{-1}$.

\begin{figure}
	\centering
	\includegraphics[width=0.43\textwidth]{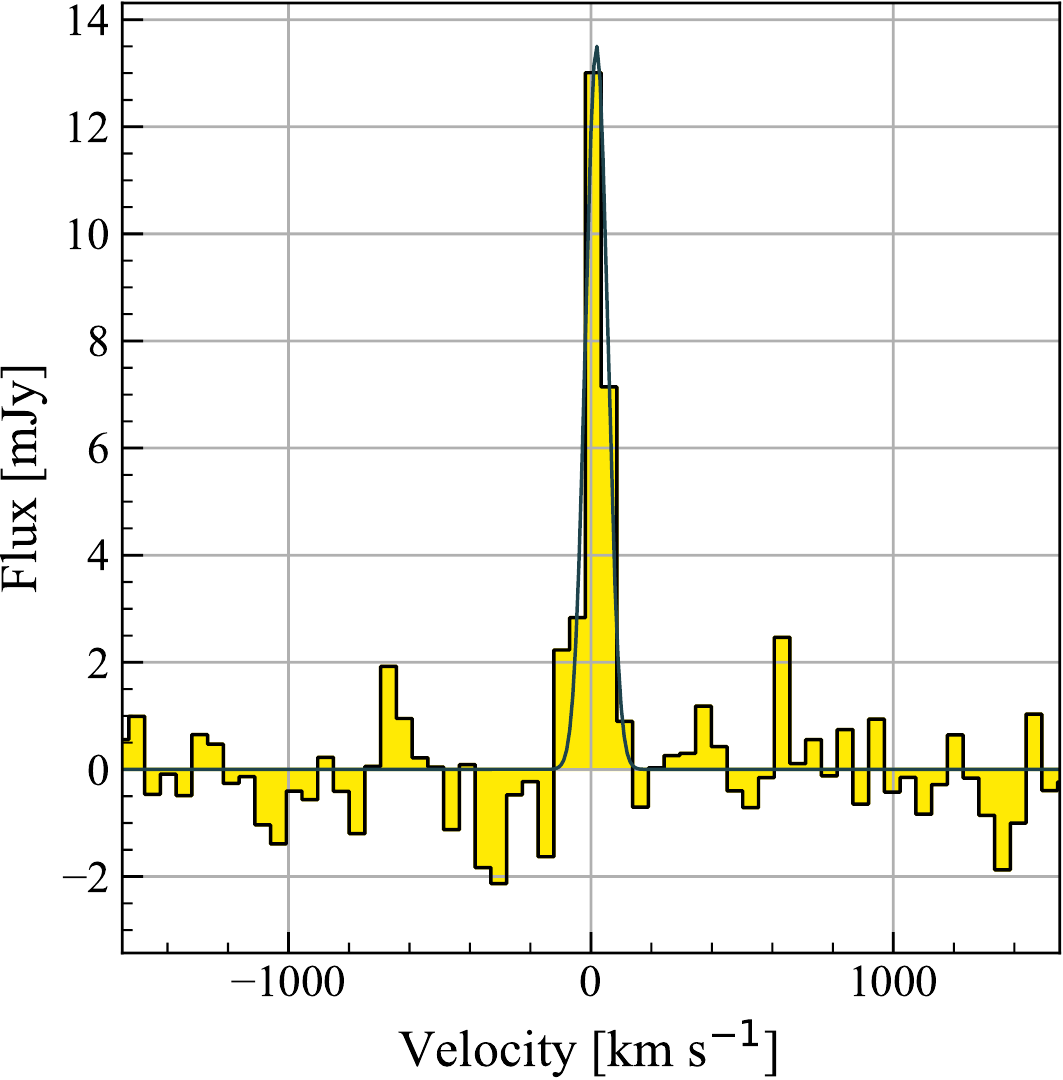}  
	\caption{NOEMA $^{12}$CO(1-0)  spectrum integrated over the H$\alpha$ blob (Totoro) and the connecting arms.  The area, 210 arcsec$^{2}$, we have used for deriving the spectrum is indicated by the dashed ellipse in Figure \ref{fig_CO_flux_vel}a. A single Gaussian fit to the line profile is overplotted. Significant $^{12}$CO(1-0) emission is detected at Totoro. }
	\label{fig_CO_spec}
\end{figure}

Figure \ref{fig_CO_spec} shows the $^{12}$CO(1-0) spectrum integrated over Totoro and the connecting arms, outlined by a dashed ellipse with an area of 210 arcsec$^{2}$ on the integrated intensity map in Figure \ref{fig_CO_flux_vel}a.
The $^{12}$CO(1-0) emission is  strongly detected in two channels  with velocities of $\sim$ 0 -- 100 km s$^{-1}$ and marginally detected   in the adjacent channels.
A single Gaussian fit to the line profile is overplotted in Figure \ref{fig_CO_spec}.
The full width  half maximum (FWHM) of the line is 86.5$\pm$7.4 km s$^{-1}$ and the line velocity relative to the galaxy velocity is 18.6$\pm$3.3 km s$^{-1}$.
The $^{12}$CO(1-0) (hereafter, CO) line properties  are summarized in Table \ref{tab_co}.

\begin{table*}
	\begin{center}
		\caption{$^{12}$CO (1-0) line properties of the H$\alpha$ blob Totoro.}
		\label{tab_co}
		\begin{tabular}{ccccc}
			\hline
			velocity &  line flux &  line luminosity & line wdith & peak flux \\
(km s$^{-1}$)  & (Jy) & (K km s$^{-1}$ pc$^{2}$) & (km s$^{-1}$)  & (mJy) \\
			\hline
			18.6$\pm$3.3 & 1.01$\pm$0.05 &(4.83$\pm$0.23)$\times$10$^{7}$ & 86.5$\pm$7.4& 13.00 $\pm$ 0.74\\ 
			\hline
		\end{tabular}
	\end{center}
\end{table*}


\section{Results and Discussion}
\label{sec_results}
In this section, we will present and discuss our results following the  scenarios  mentioned in the Introduction,  namely, Section \ref{sec_ram}: the gas  being  ram-pressure stripped from Satsuki;  Section \ref{sec_sep_gal}:  an extremely low surface brightness galaxy or ultra-diffuse galaxy;  Section \ref{sec_agn}:   gas being ejected or ionized by an AGN.
In addition, \citetalias{Lin17} did not discuss the  scenario of  cooling of the IGM, which we discuss in Section \ref{sec_cooling}.
 We will present the analysis and results of  data at a specific wavelength described in Section \ref{sec_data} when it is needed to  test a specific scenario.

\subsection{Ram-Pressure Stripping}
\label{sec_ram}

Galaxies in dense environments experience ram-pressure   \citep{Gun72}.
As pointed out in \citetalias{Lin17},  ram-pressure stripping is not expected to produce a centrally-concentrated blob, but is more likely to form clumpy structures embedded in a  jellyfish-like tail \citep[e.g.,][]{Bos16, Pog17,Bel19,Jac19}.

\begin{figure*}
	\centering
	\includegraphics[scale=0.6]{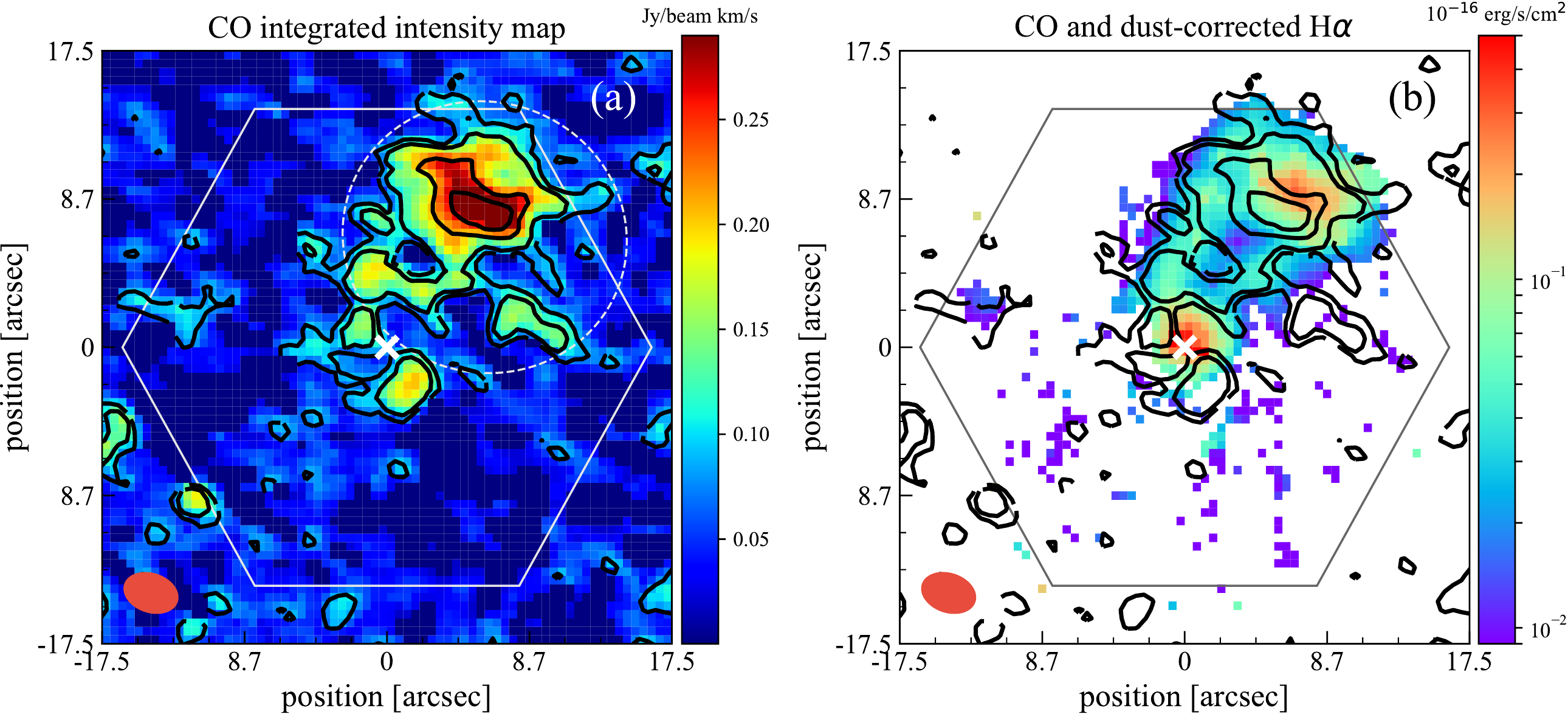}
	\includegraphics[scale=0.6]{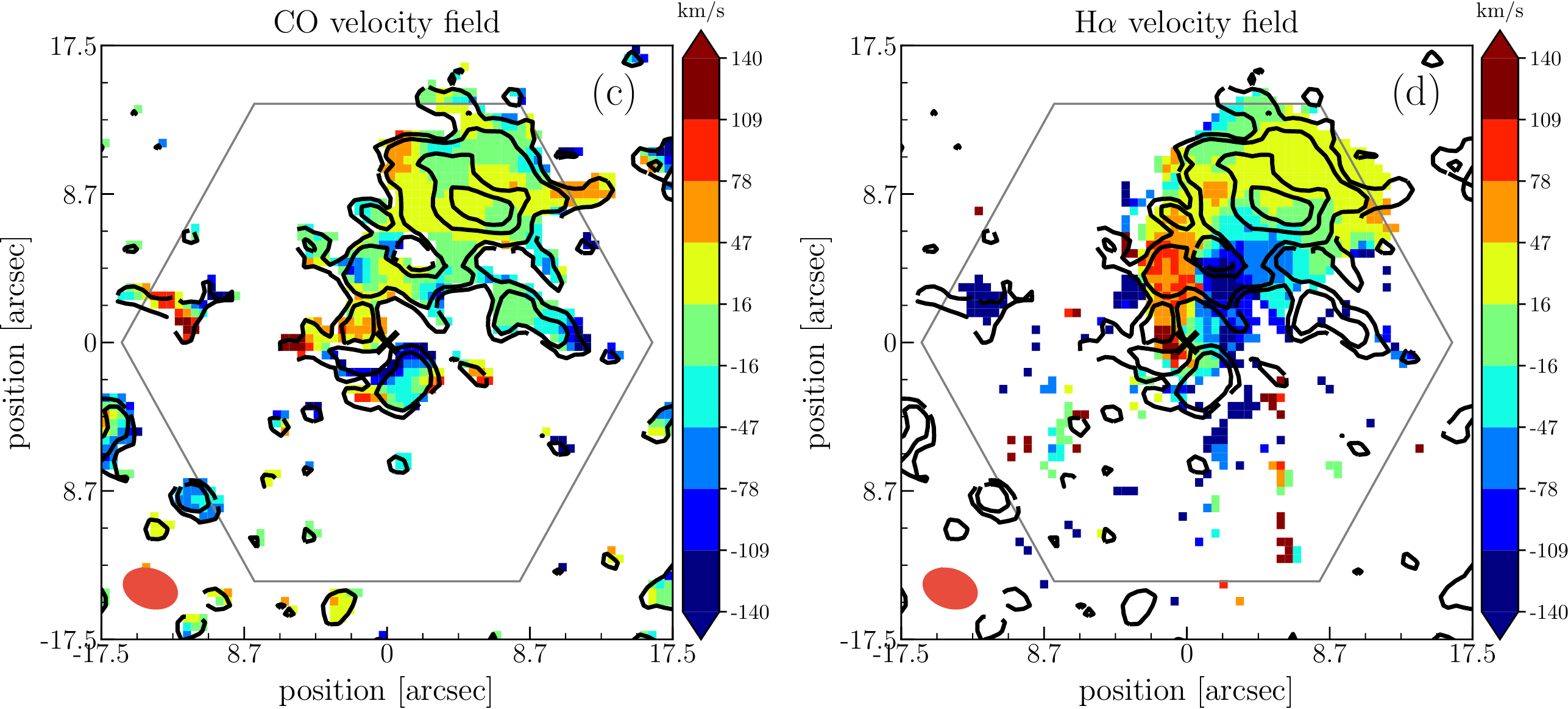}
	\caption{(a) Map of  $^{12}$CO(1-0) integrated intensity map (color and contours) with the MaNGA hexagonal FoV overlaid. The $^{12}$CO contours are in intervals of 2, 3, 5.5, and 7.5$\sigma$, where 1 $\sigma$ corresponds to 40 mJy beam$^{-1}$ km s$^{-1}$. The  nucleus of Satsuki is marked with a white cross. The synthesized beam (3.4$\arcsec$ $\times$ 2.34$\arcsec$, PA $=$ 73.3$^{\circ}$) is plotted in the bottom left.   The dashed ellipse indicates the area we have used for generating the integrated spectrum in Figure \ref{fig_CO_spec}. (b)  Similar to the panel (a), but the color map shows the H$\alpha$ emission from the MaNGA survey. (c) Velocity field of gas traced by $^{12}$CO(1-0) (color scale), with  $^{12}$CO(1-0) intensity contours overlaid. (d) Velocity field of gas traced by MaNGA H$\alpha$ (color scale). The contours are the same as in other panels.}
	\label{fig_CO_flux_vel}
\end{figure*}

The centrally peaked structure we observe in Totoro in H$\alpha$ is also seen in molecular gas.
The integrated intensity map of CO is shown in Figure \ref{fig_CO_flux_vel}a and the comparison with H$\alpha$ is displayed in Figure \ref{fig_CO_flux_vel}b 
The morphology of molecular gas generally agrees well with H$\alpha$ emission.
The CO emission of Totoro  is also  dominated by a large, centrally-concentrated structure associated with Totoro,  but the peak position of CO is offset toward the south of H$\alpha$ peak by $\sim$ 0.3$\arcsec$ ($\sim$ 200 pc).
The multi-wavelength peak coordinates of Totoro, along with other properties that will be derived and discussed in this paper are provided in Table \ref{Tab_totoro}. 
The two arm-like structures connecting the extended structure and the central region of the galaxy are also seen  in CO. 
Moreover, as somewhat expected, an early-type galaxy like Satsuki has a low molecular gas content.
 While there is a strong and compact  H$\alpha$ emission at the nucleus of Satsuki,  no CO detection is found at this position. 
Instead, two knots are  moderately detected  in the northeast and southwest of the  nucleus. Although their orientation is consistent with that of the possible past AGN jets  indicated by X-ray cavities \citep{Osu19}, deep  CO observations are required to confirm the nature of these two knots.

We cannot rule out that the H$\alpha$ and CO morphologies would appear more clumpy  if the angular resolution is improved. 
The recently reported size of  gas clumps  in  ram-pressure-stripped tails of galaxies range from several hundreds of parsec to several kpc \citep[e.g.,][]{Bel17,Lee18,Jac19,Lop20}. 
Although we are not able to resolve individual gas clumps with our $\sim$ 1.7 kpc resolution, if Totoro is intrinsically clumpy as ram-pressured stripped gas,  we should  see a sign of  clumpy sub-structure  given that the size of the object is as large as 6 -- 8 kpc in diameter.

Ram-pressure-stripped gas is known to have high velocities (several hundreds of km s$^{-1}$) and velocity dispersions ($>$ 100 km s$^{-1}$) \citep[e.g.,][]{Bel17,Con17}. 
Figure \ref{fig_CO_flux_vel}c and  \ref{fig_CO_flux_vel}d display the CO and MaNGA H$\alpha$ velocity fields in color scale.
For ease of comparison, contours of CO integrated intensity map are overplotted on both velocity fields.
There is a strong variation in the line of sight H$\alpha$ velocity at the two connecting arms, with 
redshifted gas in the left tail and blueshifted gas in the right tail. 
Such features are not observed in the  CO velocity field.
At the main blob region of Totoro,  both CO and H$\alpha$ velocity field  show no velocity gradient, little variation, and low velocity (mostly $\leq$ 60 km s$^{-1}$)  across the region, suggesting that the system is not in rotation,  unless it is perfectly face-on.  
The velocity dispersion of  H$\alpha$ gas in the region of Totoro is only $\sim$ 50 km s$^{-1}$ (\citetalias{Lin17});  the narrow CO line width also implies a low gas velocity dispersion.
Therefore, the gas kinematics of Totoro is  in conflict with that of ram-pressure stripped gas.
We should caution, however, that the velocity resolutions of our CO and H$\alpha$ map are relatively low, $\sim$ 50 and 70 km s$^{-1}$, respectively.  Further observations  are needed to probe the detailed kinematics of Totoro.

Moreover, if  ram-pressure stripping is the main origin of Totoro, the molecular-gas data imply that the cold gas is stripped almost completely from Satsuki.
However, such a scenario is disfavored by simulations of ram-pressure stripping  \citep{Ste12,Ste16}.
A galaxy can lose all of its gas only in extreme cases of ram-pressure stripping, e.g.,  galaxy encounters high ICM densities with  very high relative velocity.
In addition,   massive galaxies,  are less prone to lose their gas due to ram-pressure stripping because the existence of a massive bulge   can prevent the stripping of  gas and reduce the amount of gas being stripped.
Finally, galaxies that show ram-pressure stripping  are mostly gas-rich late-type galaxies \citep[][and series of papers by the GAs Stripping Phenomena in galaxies with MUSE, GASP, team]{Pog17}.
 For these reasons,  the CO and H$\alpha$ gas are unlikely to be moved from the center of Satsuki to the current position as a result of ram-pressure stripping. 
However, we note that this does not mean that the galaxy has not experienced any ram-pressure stripping. 
We come back to this discussion in Section  \ref{sec_cooling_env}.

\begin{table}[]
	\begin{threeparttable}
		\caption{Properties of  the offset-cooling gas Totoro.}
		\label{Tab_totoro}
		\begin{tabular}{ll}
			\hline
			\multicolumn{2}{c}{General Properties}                                            \\
			\hline
			\multirow{ 2}{*}{host}                 & VII Zw 700                                               \\
			&  (dry merger: Satsuki and Mei)  \\
			redshift\tnote{a}             & 0.0322                                                     \\
			distance to the host & $\sim$ 8 kpc                                               \\
			enviroment           & merging group                                        \\
			\hline
			\multicolumn{2}{c}{Peak Position\tnote{b}}                                                 \\
			\hline
			$^{12}$CO(1-0)                   & 17:15:22.46, $+$57:26:6.90                                    \\
			H$\alpha$\tnote{c}            & 17:15:22.40,  $+$57:26:7.83                                    \\
			X-ray                & 17:15:22.52,  $+$57:26:7.89                                    \\
			\hline
			\multicolumn{2}{c}{Luminosity}                                                    \\
			\hline
			$^{12}$CO(1-0)                   & 4.8 $\times$ 10$^{7}$ K km s pc$^{2}$                     \\
			H$\alpha$\tnote{c}            & 5.9 $\times$ 10$^{39}$ erg s$^{-1}$                        \\
			X-ray                & 4.4 $\times$ 10$^{40}$ erg s$^{-1}$                        \\
			\hline
			\multicolumn{2}{c}{Gas Mass}                                                          \\
			\hline
			cold gas (H$_{2}$)   & 2.1 $\times$ 10$^{8}$ M$_{\sun}$ ($^{12}$CO (1-0))                      \\
			& 1.9 $\times$ 10$^{8}$ M$_{\sun}$ ($A_\mathrm{V}$: cloud)   \\
			& 2.5 $\times$ 10$^{8}$ M$_{\sun}$ ($A_\mathrm{V}$: diffuse) \\
			warm gas (H$\alpha$)            & 8.2 $\times$ 10$^{4}$ M$_{\sun}$                           \\
			hot gas (X-ray)             & 1.2 $\times$ 10$^{9}$ M$_{\sun}$                           \\
			\hline
			\multicolumn{2}{c}{Other Properties}                                              \\
			\hline
			SFR\tnote{d}                 & $<$ 0.047 M$_{\sun}$ yr$^{-1}$                             \\
			cooling time         & 2.2 $\times$ 10$^{8}$ yr                                  \\
			\hline
		\end{tabular}
		\begin{tablenotes}
			\item[a] Redshift of VII Zw 700,  taken from the NASA-Sloan Atlas (http://nsatlas.org/).
			\item[b] Position of intensity peak in images, no centroid fitting is performed.
			\item[c] Based on MaNGA H$\alpha$ data.
			\item[d] Based on the assumption that all MaNGA-H$\alpha$ fluxes come from star formation.
			
		\end{tablenotes}
	\end{threeparttable}
\end{table}

\subsection{Separate Galaxy}
\label{sec_sep_gal}

In \citetalias{Lin17}, we argue that Totoro may be a separate galaxy interacting with the dry merger  (Satsuki and Mei).
We can examine this scenario by (1) searching for its underlying stellar component; (2) looking for interaction features; and  (3) comparing the molecular gas and star formation properties  of this  galaxy candidate with  other galaxy populations.

\subsubsection{Underlying  Stellar Counterpart}
\label{sec_gal_stellar}

\begin{figure*}
	\centering
	\includegraphics[width=0.9\textwidth]{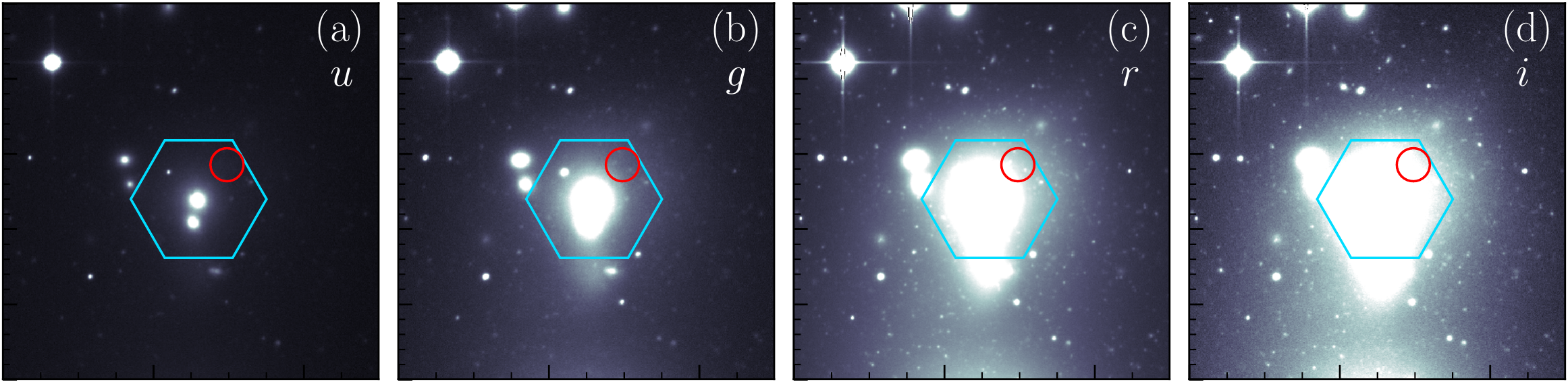}  
	\caption{CFHT $u$-, $g$-, $r$-, and $i$-band images (from left to right). The first three bands  combine archival data downloaded from the CADC server and the data taken in 2015 summer (see \citetalias{Lin17} for the details). The $u$-band data was taken later in 2017. The cyan hexagon and  red circle mark the region of MaNGA FoV and the H$\alpha$ blob Totoro.} 
	\label{fig_CFHT_all}
\end{figure*}

The CFHT $u$-, $g$-,  $r$, and $i$-band images are presented in Figure  \ref{fig_CFHT_all}a --  \ref{fig_CFHT_all}c, respectively.
 There are extended stellar halos surrounding the two galaxies, but  we find  no apparent optical counterpart directly from the images at the position of Totoro.
The limiting magnitude\footnote{The $gri$ limiting surface brightnesses quoted here are different from that given in \citetalias{Lin17}. This is because the former is scaled from the limiting magnitudes using 1$\arcsec$ aperture size  while the latter was measured using an aperture size corresponding to one arcsec$^{2}$ area, which however is more affected by the Point Spread Function (PSF) because of the small aperture.} and the surface brightness of Totoro are listed in Table \ref{tab_cfht_mag}. Due to the absence of  an optical counterpart at the position of Totoro,  only upper limits can be placed on  surface brightnesses (i.e., limiting surface brightness of the observations).

In \citetalias{Lin17}, we used a multi-component \texttt{GALFIT}  \citep{Pen10} model for searching for a stellar component of Totoro from a $g$-band image and   found no sub-structure that is responsible for the blob. 
However,  the residual of this complex parametric model still showed significant fluctuations that could hinder the detection of small scale sub-structure (see Figure 6 in \citetalias{Lin17}). 
The on-going interaction of the main system creates an extended stellar envelope and asymmetric structures that are challenging to the model.
In this paper, we use three different approaches to further search for a potential stellar counterpart to Totoro from   multiple-band ( $g$, $r$, and $i$) images.
All three methods are commonly used in literature for   background subtraction and   for both compact and extended source  detection.

\begin{table}[]
	\begin{center}
	\caption{The 5$\sigma$ limiting magnitudes of CFHT $u$-, $g$-, $r$-, and $i$-band images and the  5$\sigma$ upper limit of the surface brightness of Totoro at each band.}
\label{tab_cfht_mag}
	\begin{tabular}{ccc}
		\hline
		band & limiting magnitude  & surface brightness of Totoro \\
		& (1$\arcsec$ radius) [mag]  & (upper limit) [mag arcsec$^{-2}$]  \\
		\hline
		$u$ & 26.3                  & 27.54                           \\
		$g$  & 25.7                 & 26.94                     \\
		$r$ &    26.2                &      27.44                        \\
		$i$ &      25.2              &           26.44                  \\
		\hline
	\end{tabular}
	\end{center}
\end{table}

Firstly, we use the Python photometry tool \texttt{sep} \citep{Bar16}\footnote{https://github.com/kbarbary/sep} to generate a ``background'' model of the image with small background box size (5 pixels). 
The tool \texttt{sep} uses the same background algorithm as in \texttt{SExtractor} \citep{Ber96}. 
We then subtract the ``background'' model from the image to increase the contrast around the main galaxies. 
This method has been used to  separate galaxy stellar halos and  the light from adjacent (background) objects \citep[e.g.,][]{Hua18,Rub18}, and  to identify faint, extended emission such as tidal tails \citep[e.g.,][]{Man19}.

As a second approach, we subtract a blurred version of the image from the original one. 
The blurred version of the image is created by convolving the image with a circular Gaussian kernel ($\sigma = 4$ pixels).
This procedure is part of the unsharp masking method in digital image processing\footnote{The original unsharp mask technique re-scales the residual and add it back to the original image.} and can also increase the contrast of the image. 
Compared to the first approach, the Gaussian convolution makes it more sensitive to a low threshold feature with a sharp edge. 
 This method has  been commonly used to identity HII regions in galaxies (e.g., \citealt{Rah11}; Pan et al.   in preparation)   and to detect faint  embedded  spiral and bar features in  early-type galaxies \citep[e.g.,][]{Bar02,Kim12}. The method has been applied to IFS data for the  later purpose as well by \cite{Gom16}.

For our third method, we perform isophotal fitting using the \texttt{Ellipse} task in \texttt{IRAF} \citep{Tod86,Tod93}. 
We first mask out all detected objects other than the main galaxy using \texttt{sep}. 
Then we run \texttt{Ellipse} on the masked images, allowing the centroid and the shape of the isophote to vary. 
Using the resulting isophotal parameters, we create the corresponding 2-D model using the \texttt{bmodel} task and subtract it from the input image. 
Compared to the \texttt{GALFIT} parametric model, the \texttt{Ellipse} one does not depend on the choice of model component and typically leads to smoother residuals. 
This approach has been routinely used  to determine  morphology of elliptical and lenticular galaxies  \citep[e.g.,][]{Hao06,Oh17} and to search for low-surface-brightness tidal features in nearby galaxies \citep[e.g.,][]{Tal09,Gu13}. 

\begin{figure}
	\includegraphics[width=0.45\textwidth]{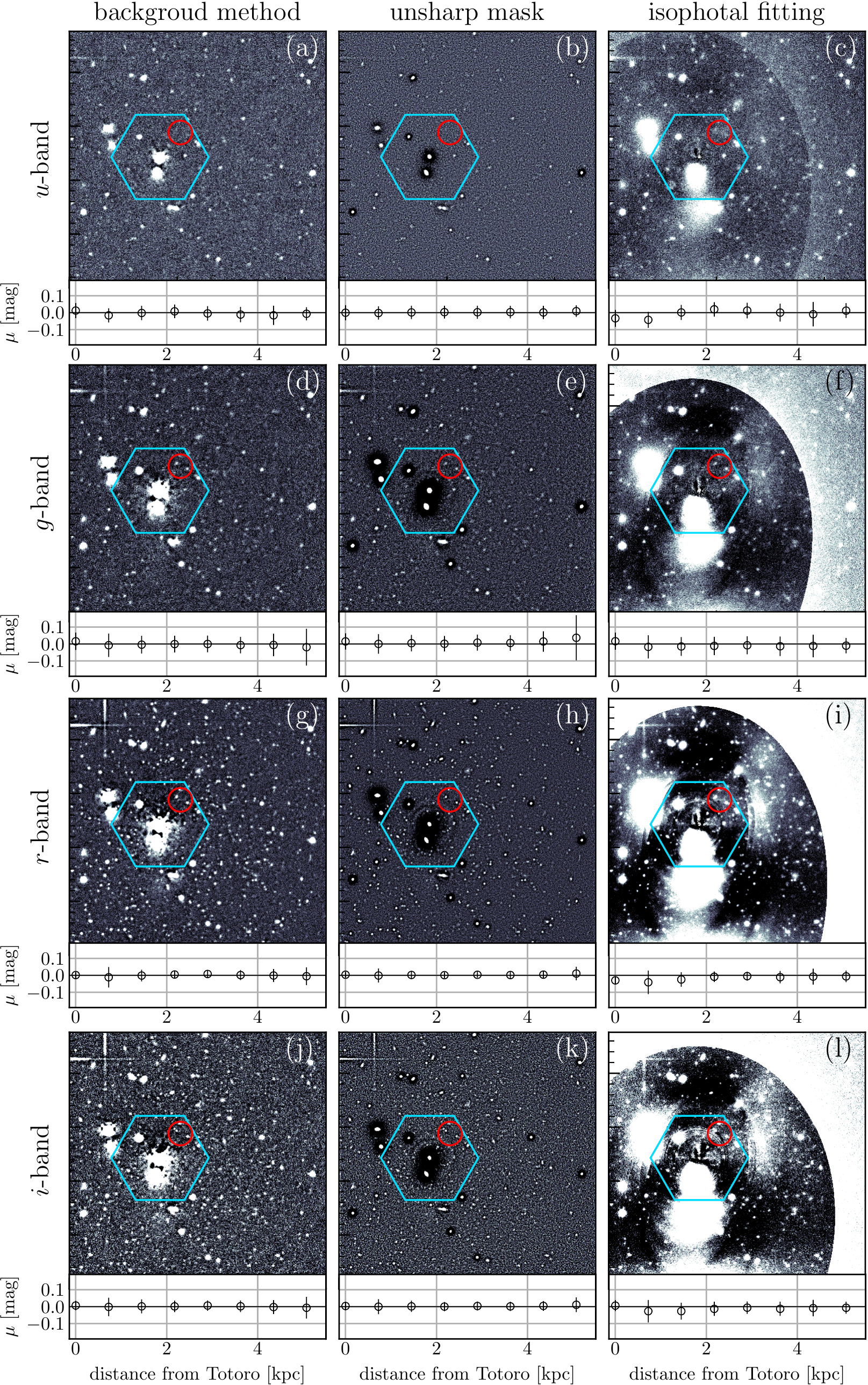}  
	\caption{The  residual images and annular residual profiles   after subtracting the model images of the dry merger  (Satsuki and Mei).  The  annular  profiles  are centered on the H$\alpha$ peak of Totoro. Three different approaches (from left to right: background method, unsharp mask, and isophotal fitting) are used to  search for distinctive stellar counterpart and star formation of Totoro on the  $u$-, $g$-, $r$- and $i$-band images (from top to bottom). The cyan hexagon and the red circle mark the regions of MaNGA FoV and Totoro, respectively.  We do not observe a significant, extended stellar component around with Totoro in any residual maps and profiles. There are two point sources around the position of Totoro,  they are likely background sources (see text for the details). }
	\label{fig_lightfitting}
\end{figure}

The  residual maps of these methods are shown in Figure \ref{fig_lightfitting}.  
From left to right the panels correspond to the analysis of the background, unsharp mask, and isophotal fitting methods; from top to bottom we show the residual maps and annular residual profiles  of $u$-, $g$-, $r$- and $i$-band, respectively ($u$-band results will be discussed later in this section).
The  annular  profiles  are centered on the H$\alpha$ peak of Totoro.
We also explore different parameters adopted in these procedures (e.g., the size of the background box or the convolution kernel). 
The choice of these parameters within a reasonable range does not affect the results.

Using these methods, we detect a large number of  unresolved (point-like) sources on the $g$-, $r$- and $i$-band images, presumably a combination of globular clusters of the main system and  background galaxies.
 The residual profiles are generally flat, fluctuating around the zero value.
	This is true for all $gri$ bands and methods, suggesting that there is no sign of a distinctive stellar counterpart at the position of Totoro.
Therefore,  our new analyses confirm the previous result by using \texttt{GALFIT}  in \citetalias{Lin17}.
There are two unresolved sources in the Totoro area (red circle in Figure \ref{fig_lightfitting}), but we do not notice any increase or decrease of number density around the Totoro area.  
The two sources are  associated with neither H$\alpha$ nor CO peak.
We will come back to the nature of these two unresolved sources later.
On the  residual maps from the unsharp mask method, we uncover a pair of ``ripple''-like features close to the center of the main galaxy. 
These sub-structures remain the same when we vary the convolution kernel used, and they resemble the structure we saw on the \texttt{GALFIT} residual maps in \citetalias{Lin17}. 
Such ``ripples'' often relate to recent galaxy interaction or the presence of dust \citep[e.g.,][]{Col01,Kim12,Duc15,Bil16}. 
However, it is unclear whether there is any connection between them and Totoro.


Since there is no optical ($g$, $r$, and $i$) continuum counterpart found in  Totoro, it is expected to be composed mostly of young stars if it is indeed a  galaxy.
In \citetalias{Lin17}, we use the excitation state of optical lines to constrain the presence of young stars. However, the result is  method (diagnostic diagram) dependent.
The $u$-band luminosity  of  a  galaxy  is  dominated  by young  stars  of  ages $<$ 1 Gyr, therefore it  is more sensitive to any recent  star formation than any of the other broad band luminosities available \citep{Mou06,Pre09,Zho17}, and is a more straightforward probe than optical emission line diagnostics.   

Similar to the $gri$ bands, the $u$-band emission in the MaNGA hexagonal FoV  is dominated by the dry merger  (Satsuki and Mei) as shown in Figure  \ref{fig_CFHT_all}, but the $u$-band data  is less affected by the large stellar halos associated with the dry merger  (Satsuki and Mei) than the redder bands, and therefore serves as a better probe for underlying stellar component.
 The limiting magnitudes and surface brightness of the $u$-band image are 26.3 mag and 27.54 mag arcsec$^{-2}$, respectively (Table \ref{tab_cfht_mag}). 
The  point source  in the upper-left corner of the dry merger  (Satsuki and Mei) is a foreground star according to the MaNGA spectrum.


Although  visually there is no distinguishable $u$-band feature (i.e.,  recent star formation) associated with Totoro, to ensure that the  $u$-band counterpart  of Totoro  is not embedded within the light of the dry merger  (Satsuki and Mei), we subtract the photometric models for the merging system from the $u$-band image using the three different methods mentioned above.
The residual maps and profiles are shown in the top row of Figure \ref{fig_lightfitting}, respectively.
The residual maps and profiles are not perfectly smooth, but we find no obvious evidence for a distinctive, extended  $u$-band counterpart at Totoro.
The two point sources seen in $g$-, $r$-, and $i$-band residual maps are also seen in the $u$-band residual image of the isophotal fitting (and  marginally seen  in the background method as well).
This confirms the results in \citetalias{Lin17} that star formation alone can not explain the excitation state of Totoro.

\begin{figure}[!ht]
	\begin{center}
		\subfigure[]{\label{fig_PS_S}\includegraphics[scale=0.4]{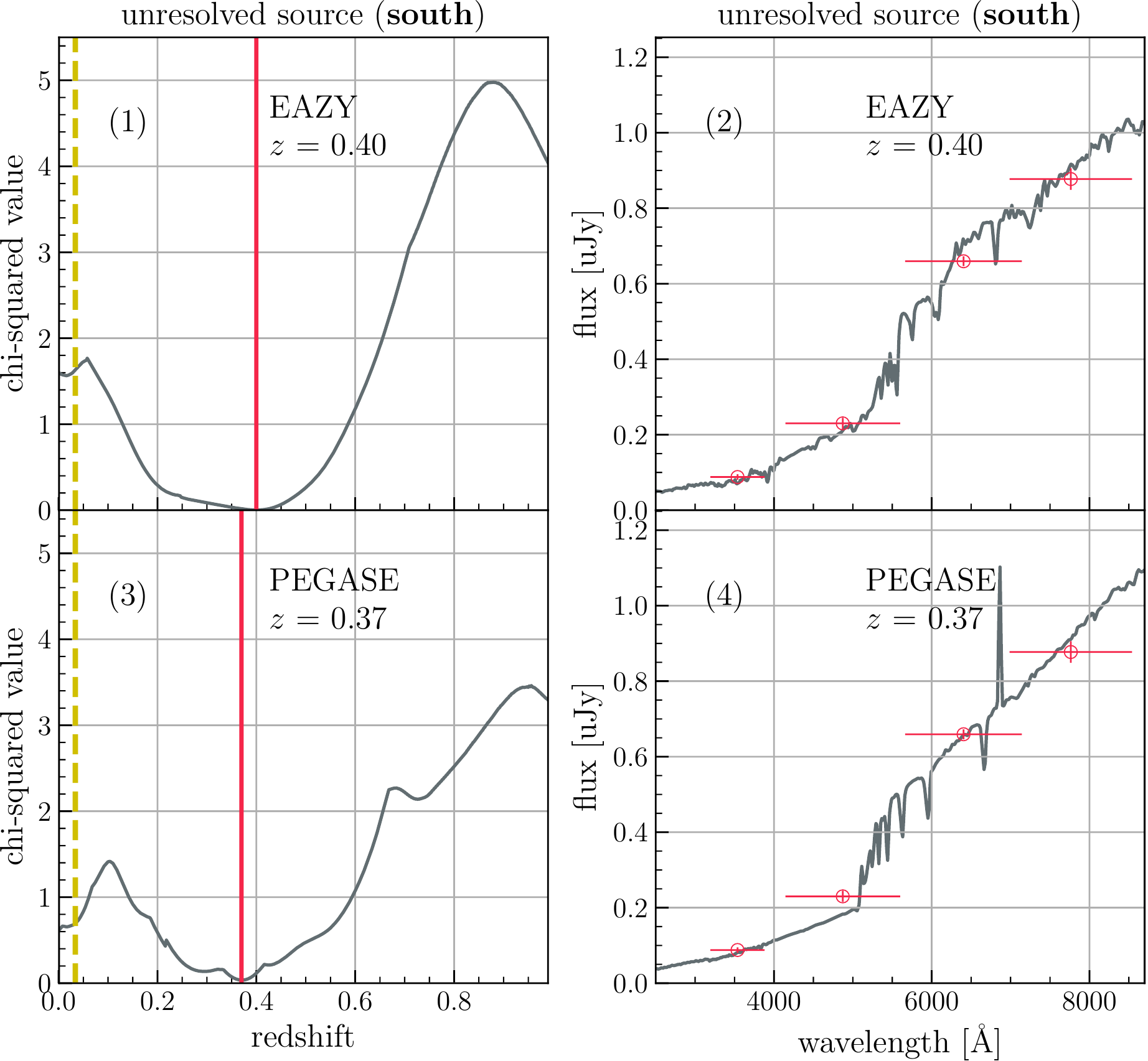}}
		\subfigure[]{\label{fig_PS_N}\includegraphics[scale=0.4]{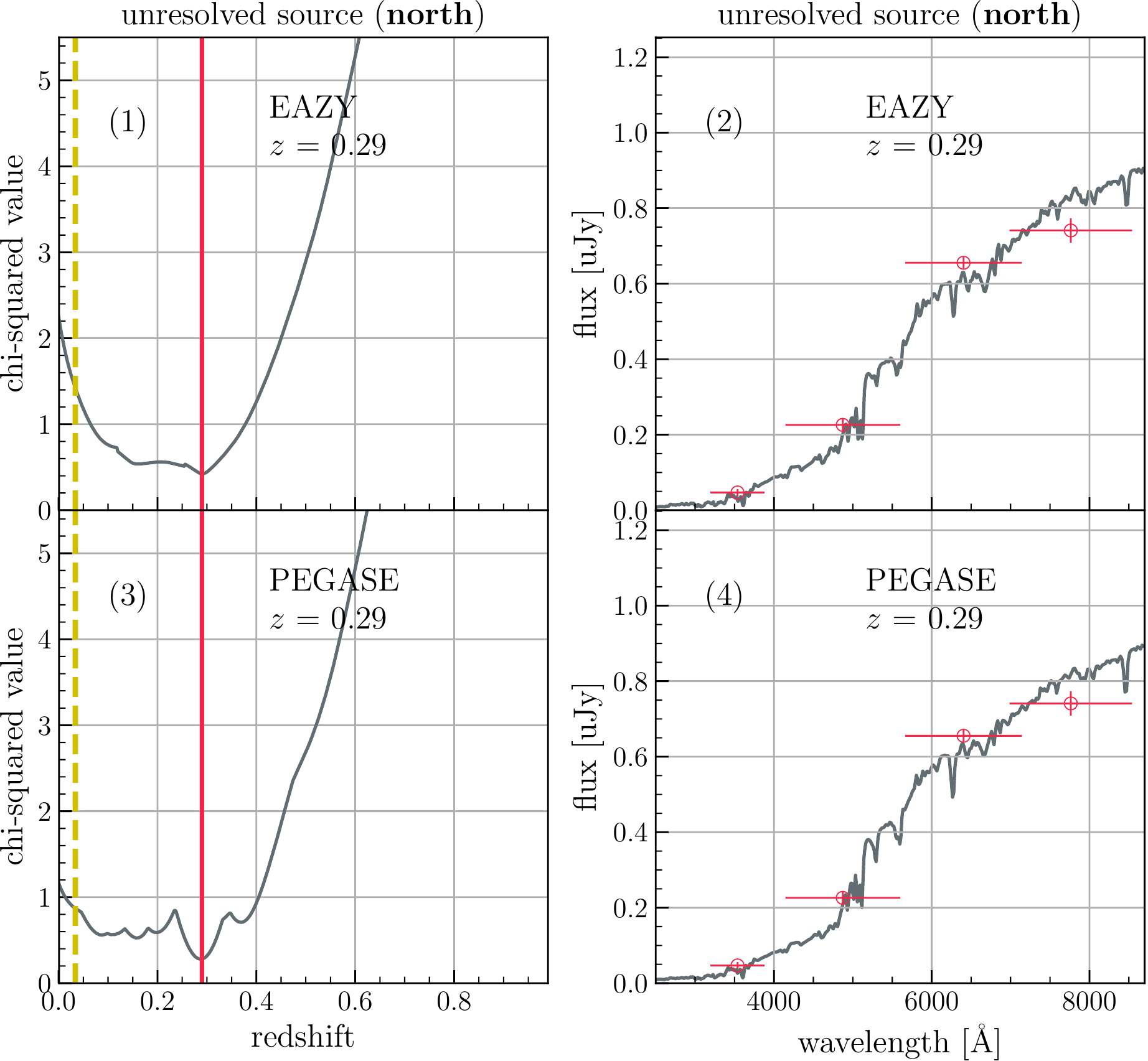}}
	\end{center}
	\vspace{-15pt}
	\caption{The SED-fitting results of the two unresolved (point-like) sources appearing in the residual images after subtracting the model images of the dry merger (Satsuki and Mei). The results of the southern and northern sources are presented in panel (a) and (b), respectively. For each panel, the sub-panels  (1) and (2) show  the chi-squared values of the fit for a given redshift using  the EAZY template and  the best-fit SED   with observed fluxes at   $i$-, $r$-, $g$- and $u$-band  overlaid as red circles, respectively.  The corresponding plots using the   P\'{E}GASE template are shown in  sub-panels  (3) and (4). The red solid vertical lines in sub-panels (1) and (3) mark the redshift of minimum  chi-squared value. The redshift of our target  is indicated by yellow dashed lines.
	}
	\label{fig_sed}
\end{figure}

To gain some insight into the nature of the two unresolved objects in the Totoro area, their photometric redshifts are determined using the EAZY \citep{Br08} and P\'{E}GASE 2.0  \citep{fr1997} template fitting the   aperture magnitudes measured from the isophotal fitting residual images  using GAIA (Graphical Astronomy and Image Analysis Tool).
The default EAZY template is generated from the  P\'{E}GASE 2.0 models 
using the \citet{br2007} algorithm and then calibrated using 
semi-analytic models,  plus an additional young and dusty template.
The P\'{E}GASE 2.0 template is a library including
$\sim$3000 models with a variety of star formation histories 
and with ages between 1 Myr and 20 Gyr, as described in detail
in \cite{grazian2006}.
Figure  \ref{fig_sed} shows the  chi-squared of the fit for a given redshift using  the EAZY and P\'{E}GASE  templates (left column of each panel) and  the best-fit SED   with observed fluxes at   $i$-, $r$-, $g$- and $u$-band  overlaid as red circles  (right column of each panel).
The results based on the EAZY and P\'{E}GASE  templates  are presented in the top and bottom rows respectively.
The minimum  chi-squared value indicates that the southern  source (17$^\mathrm{h}$15$^\mathrm{m}$22.111$^\mathrm{s}$, +57$^{\circ}$26$\arcmin$5.628$\arcsec$) is a background galaxy at $z$ $\sim$ 0.40 and 0.37  using  the EAZY and  P\'{E}GASE  templates, respectively (Figure \ref{fig_PS_S}).
The redshift of our target is marked by a yellow dashed  line in the figures.  We notice a second minimum at $z$ $<$ 0.03, close to the redshift of our target. However, at such low redshifts, the source would  be resolved, not point like.
The most plausible redshift  of the northern source  (17$^\mathrm{h}$15$^\mathrm{m}$22.573$^\mathrm{s}$, +57$^{\circ}$26$\arcmin$7.659$\arcsec$) is  0.29 according to both templates (Figure \ref{fig_PS_N}), but  we cannot rule out other possibilities of $z$ $<$ 0.4,  in particular $z$ $\sim$ 0.15,    due to the shallower basin-shaped chi-square distribution. Nonetheless, the redshift of Totoro (yellow dashed  line) is not associated with any local minimum of the chi-square values. We find poorer fits when using stellar templates, providing further support that the source is not a nearby object.
The best-fitted SEDs from  P\'{E}GASE  are exported to SED-fitting code \texttt{New-Hyperz}\footnote{\url{http://userpages.irap.omp.eu/~rpello/newhyperz/}} in order to derive the stellar mass ($M_{\ast}$) and star formation rate (SFR) of these two objects. 
The results suggest that the southern and northern sources are   $\sim$3 $\times$ 10$^{8}$ and  $\sim$6 $\times$ 10$^{8}$ M$_{\sun}$ and their specific SFR (sSFR $=$ SFR/$M_{\ast}$) are  $\leq$ 10$^{-11}$ yr$^{-1}$.
We also estimate their $M_{\ast}$  assuming they are at the same redshift as Totoro ($z$ $=$ 0.03), and yield $\sim$ 10$^{6}$ and $\sim$ 10$^{7}$ M$_{\sun}$ for the southern and northern sources, respectively.  
Given their point-like  morphologies ($<$ 1 kpc$^{2}$), we expect to see distinctive   stellar mass surface density ($\Sigma_{\ast}$) distributions in the MaNGA data  if they are associated with Totoro. However, we find no such features.
It should be noted that  the above stellar masses   may be subjected to non-negligible uncertainties due to the lack of (near-) infrared measurements.


We  use  web interface Marvin\footnote{https://dr15.sdss.org/marvin},  a tool to visualise and analyse MaNGA data \citep{Che19}, to search for   any signatures in the spectra of these two sources by  redshifting strong optical emission lines (e.g., H$\alpha$) based on the derived  photometric redshifts. 
However, the emission lines are too faint to be seen by MaNGA. 
Given their likely high redshift and their positions being offset from the centroid of the H$\alpha$ blob, these two point sources are  unlikely to be associated with Totoro.

\subsubsection{Interaction Features}
\label{sec_gal_ion}

\begin{figure*}
	\centering
	\includegraphics[width=0.99\textwidth]{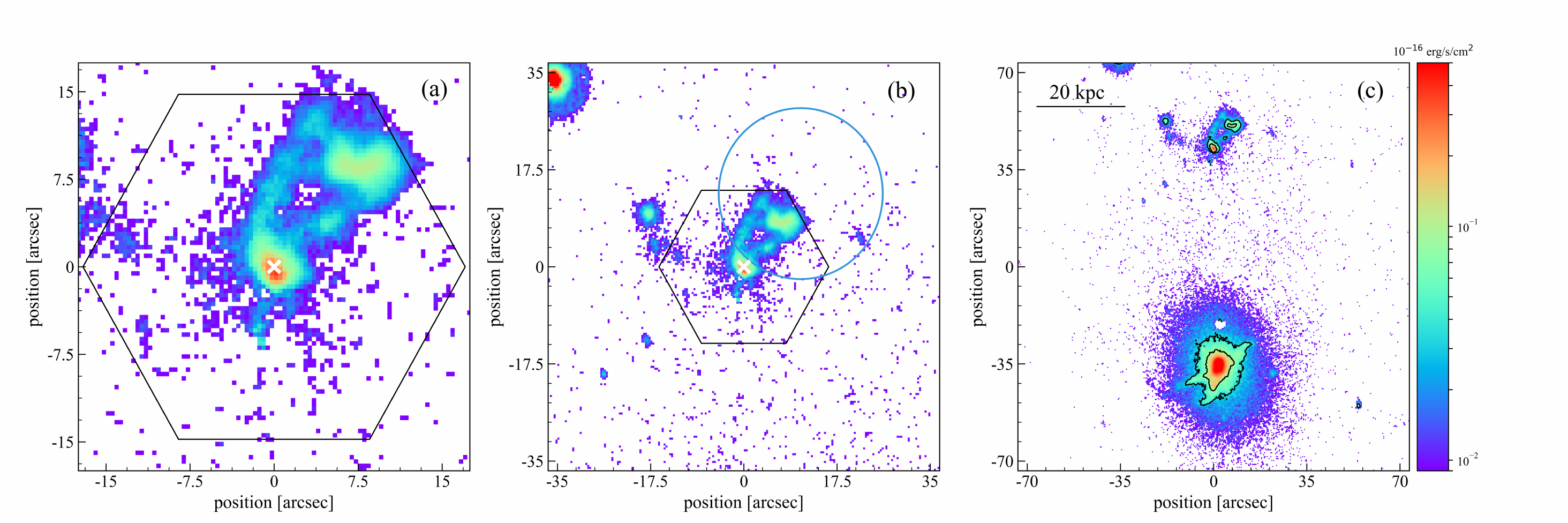}  
	\caption{New, wide-field  narrow-band H$\alpha$ image taken from the SAO RAS 6-m telescope.    (a) A zoom-in to the region of the MaNGA observation with  the MaNGA hexagonal bundle FoV overlaid.  The color scale is  the same as in Figure \ref{fig_Halpha}. The nucleus of Satsuki is  marked by a cross. (b) A zoom-out view of the  region of interest. It is clear that there is no hint of tidal feature extending beyond the MaNGA FoV.  Quantitatively, the total H$\alpha$ luminosity in the blue circle (15$\arcsec$ or $\sim$ 9 kpc in radius) is only $\sim$ 10\% higher than the luminosity of Totoro measured by MaNGA, therefore the possible missing flux that is associated with Totoro  due to the small  MaNGA FoV is at most 10\%.  (c) The   H$\alpha$ image of NGC 6338 and our target. The two contours correspond to 0.035 and 0.09 $\times$ 10$^{-16}$ erg s$^{-1}$ cm$^{-2}$, respectively.}
	\label{fig_Ha_large}
\end{figure*}

If Totoro is indeed a separate galaxy, it  may possess another tidal tail on the other side of the blob (i.e., as opposed to the ones that connect the Totoro and Satsuki), but such tail(s) would not be seen by MaNGA due to the limited FoV.
Figure \ref{fig_Ha_large} shows the new  wide-field H$\alpha$ image taken from the SAO RAS 6-m telescope. 
The new H$\alpha$ map has a sensitivity comparable to that of MaNGA, but the FoV is $\sim$ 10 times larger.
For comparison, Figure \ref{fig_Ha_large}a  zooms in to the region of the MaNGA hexagonal FoV.
The new H$\alpha$ image globally resembles that of the MaNGA H$\alpha$ in Figure \ref{fig_Halpha}, but shows finer structures  thanks to higher spatial resolution.

A zoom-out view of the  region of interest is displayed in Figure \ref{fig_Ha_large}b.
 There are several H$\alpha$ knots   beyond the MaNGA FoV towards the east. These are background galaxies or foreground stars.
It is clear that there is no hint of H$\alpha$ emission extending beyond the MaNGA FoV   at the opposite side Totoro.
Quantitatively, we can estimate the possible missing flux of MaNGA by comparing the H$\alpha$ flux related with Totoro from the two observations.
The total H$\alpha$ luminosity of Totoro  and the surrounding $\sim$15$\arcsec$ ($\sim$ 9 kpc) region outside of the MaNGA FoV (blue circle in Figure  \ref{fig_Ha_large}b) is 6.5 $\times$ 10$^{39}$ erg s$^{-1}$.
The total  H$\alpha$ luminosity of Totoro measured by MaNGA is 5.9 $\times$ 10$^{39}$ erg s$^{-1}$.
Therefore the possible missing flux of Totoro due to limited MaNGA FoV  is no larger than 10\%. 
Accordingly, the scenario of a separate galaxy  appears less likely unless the tidal tail(s)   develop at  only one side of a galaxy.
Such cases are relatively rare, though not impossible  (e.g., Arp 173, Arp188, and Arp 273), depending on the stage of the interaction and the projected orientation on the sky \citep[e.g.,][]{Mih04,Str12}. 
However, the low gas velocity provides additional support against a merger scenario (Figure \ref{fig_CO_flux_vel}c and \ref{fig_CO_flux_vel}d).


On the other hand, it is possible that Totoro is a  completely disrupted dwarf galaxy.
The  averaged surface brightness of Totoro has an upper limit of $\sim$ 27 mag arcsec$^{-2}$ (Table \ref{tab_cfht_mag}).
It would be classified as a dwarf low-surface-brightness galaxy  (LSB; $>$ 23 mag arcsec$^{-2}$) if  it is a galaxy.
In the next section, we compare the star formation and cold gas properties of Totoro with other galaxy populations, including LSBs.

As a side note, the nearby galaxy NGC 6338 is encompassed by the wide-field H$\alpha$ observation.  
Figure \ref{fig_Ha_large}c shows the H$\alpha$ image of  NGC 6338.
There is no strong  filaments or bridge linking NGC 6338 and VII Zw 700, presumably because  the two are very separate entities (with projected separation of $\sim$ 42 kpc  and  $\sim$ 1400 km s$^{-1}$ difference in line of sight velocity), but there  is a bit more H$\alpha$ emission between galaxies than  the rest of the regions in the plotted area.
The H$\alpha$ emission of NGC 6338 is  characteristic of three previously-reported filaments in the southeast and northwest quadrants. 
The H$\alpha$ intensity contours of 0.035 and 0.09 $\times$ 10$^{-16}$ erg s$^{-1}$ cm$^{-2}$ are overplotted to highlight the asymmetric filaments.
We also refer the reader to \citet{Mar04} and \citet{Osu19} for higher-resolution H$\alpha$ images of NGC 6338 and \citet{Gom16} for optical IFS data analysis of NGC 6338.

\subsubsection{$M_\mathrm{H_{2}}$-SFR  Relation of Galaxies}
\label{gas_sfr_mh2}

\begin{figure}
\centering
	\includegraphics[width=0.43\textwidth]{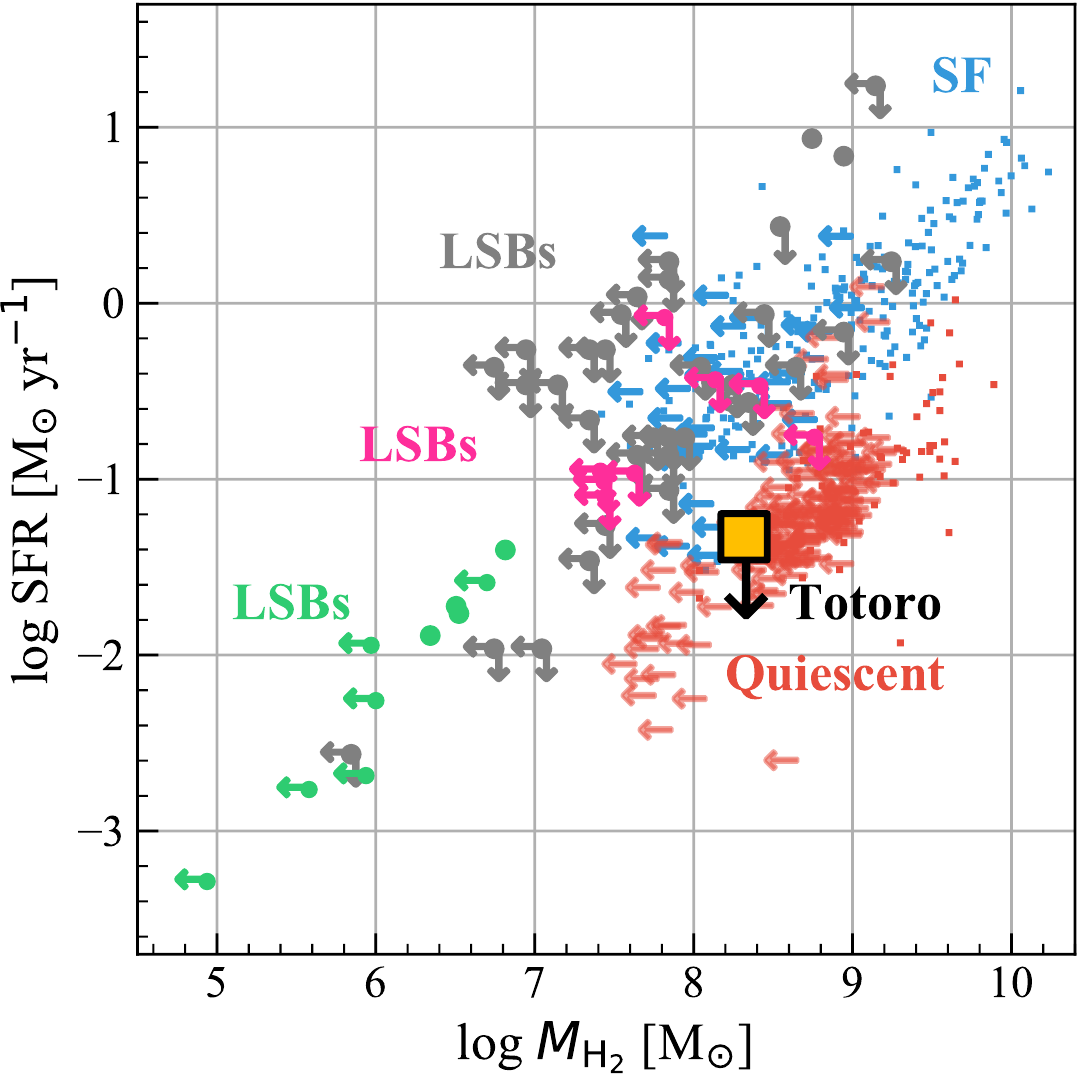}  
	\caption{Star formation rate versus H$_{2}$ mass. The orange square is Totoro from this work. The gray,  green, and magenta symbols are LSBs measurements from \cite{One03}, \cite{Mat05}, and \cite{Cao17}, respectively. Objects with solid detection in CO lines are shown with squares.  An arrow indicates that only an upper limit was found. Blue and red squares are nearby star-forming (sSFR $>$ 10$^{-11}$ yr$^{-1}$) and quiescent (sSFR $<$ 10$^{-11}$ yr$^{-1}$) galaxies taken from the xCOLD GASS survey \citep{Sai17}. In all cases a conversion factor of 2.0 $\times$ 10$^{20}$ cm$^{-2}$ (K km s$^{-1}$)$^{-1}$ \citep[][the uncertainty of the  conversion factor is $\sim$ 30\%]{Bol13} is  used to allow ready comparison between the studies. }
	\label{fig_CO_ks}
\end{figure}

Another way to constrain the origin of Totoro using data in hand  is to look into the question of whether Totoro shares  similar gas and star formation properties with nearby galaxies. 
This could not be concluded in \citetalias{Lin17} due to the lack of cold gas data.
Figure \ref{fig_CO_ks} compares the star formation rate (SFR) and molecular gas mass ($M_\mathrm{H_{2}}$)  of Totoro  with other galaxy populations.
The plot resembles the Kennicutt-Schmidt relation \citep{Ken89}  assuming that the molecular gas and star-forming regions coexist. 
The galaxy data include nearby LSBs \citep{One03,Mat05,Cao17},  nearby star-forming (sSFR $>$ 10$^{-11}$ yr$^{-1}$) and quiescent  (sSFR $<$ 10$^{-11}$ yr$^{-1}$) galaxies \citep{Sai17}.

The total  H$\alpha$ luminosity of Totoro (including the connecting arms; $\sim$48.1 kpc$^{2}$ area in total)  is converted to SFR using the  calibration of 
\begin{equation}
\frac{\mathrm{SFR}}{[\mathrm{M_{\odot}\, yr^{-1}}]}=7.9\, \times\, 10^{42}\,\frac{L_{\mathrm{H\alpha }}}{[\mathrm{erg\, s^{-1}}]}
\end{equation}
 \citep{Ken98}, where $L_\mathrm{H\alpha}$ is H$\alpha$ luminosity.
This yields a  SFR of 0.047 M$_{\sun}$ yr$^{-1}$.
 Here we assume all of the H$\alpha$  results from star formation, but this  is unlikely to be the case as BPT diagnostics  indicate a composite nature, LI(N)ER-HII mix excitation, for Totoro (\citetalias{Lin17}).
Therefore the derived SFR is an upper limit.

The total molecular gas mass of Totoro  traced by CO is computed using 
\begin{multline}
\frac{M\mathrm{_{H_{2}}}}{[\mathrm{M_{\odot}}]}=1.05\, \times\, 10^{4}\, \frac{X_{\mathrm{CO}}}{[2\, \times\, 10^{20}\, \mathrm{cm^{-2}(K\, km\, s^{-1})^{-1}}]}\, \\ \frac{S_{\mathrm{CO}}\Delta v }{[\mathrm{Jy\, km\, s^{-1}}]}\frac{D_\mathrm{L}^{2}}{[\mathrm{Mpc}]}\, (1+z)^{-1}
\end{multline}
,where $X_\mathrm{CO}$, $S\mathrm{_{CO}}\Delta\nu$, $D_{\mathrm{L}}$ are CO-to-H$_{2}$ conversion factor,  integrated line flux density, and luminosity distance, respectively \citep{Bol13}.
We adopt a Galactic $X_\mathrm{CO}$ of 2 $\times$ 10$^{20}$ cm$^{-2}$ (K km s$^{-1}$)$^{-1}$ \citep[][]{Bol13}.
The derived $M_\mathrm{H_{2}}$\footnote{We should note that the  value of $M_\mathrm{H_{2}}$ depends on the  choice of $X_\mathrm{CO}$.
	We may overestimate  $M_\mathrm{H_{2}}$ by $\sim$ 30\%  to a factor of a few if the metallicity of Totoro is indeed higher than the solar  metallicity by $\sim$ 0.3 dex as discussed in the Introduction \citep{Bol13}. 
	\cite{Van17} use  the optical thin  $^{13}$CO(3-2) emission line to estimate $X_\mathrm{CO}$ of the BCG of the cooling-core cluster  RXJ0821$+$0752,  finding a $X_\mathrm{CO}$ of  a factor of two lower than the Galactic (our adopted) value.  
	However,  this  is within the object-to-object scatter from extragalactic sources, and based on several assumptions such as  isotopic abundance ratio and excitation line ratios.
	Statistical analysis is necessary in order to constrain the $X_\mathrm{CO}$ in BCGs.} of Totoro is (2.1$\pm$0.1) $\times$ 10$^{8}$ M$_{\sun}$. 
	Note that the uncertainty due to different  methodologies to derived the Galactic $X_\mathrm{CO}$  is   $\sim$ 30\% \citep{Bol13}.

In addition to CO, the gaseous column density $N_\mathrm{H_{2}}$ (and  $M_\mathrm{H_{2}}$) also follows  from the amount of extinction $A_\mathrm{V}$, which can be derived from the MaNGA H$\alpha$ and H$\beta$ data.
For reference, the mean $A_\mathrm{V}$ across Totoro is $\sim$ 0.5 mag.
The conversion  from  extinction  to  H$_{2}$  column  density  depends on the medium  \citep{Boh78,Eva09}:
\begin{align}
\label{eq_n2av_mc}
\frac{N(\mathrm{H_{2}})}{[\mathrm{cm^{-2}}]}&=6.9\, \times\, 10^{20}\, \frac{A_\mathrm{v}}{[\mathrm{mag}]}\, \mathrm{(molecular\,cloud)} \\
\label{eq_n2av_diffuse}
&= 9.4\, \times\, 10^{20}\, \frac{A_\mathrm{v}}{[\mathrm{mag}]}\, \mathrm{(diffuse\,ISM)}.
\end{align}
These yield a $M_\mathrm{H_{2}}$ of $\sim$ 1.9 $\times$ 10$^{8}$ and $\sim$ 2.5 $\times$ 10$^{8}$  M$_{\odot}$ assuming molecular-cloud- (Equation \ref{eq_n2av_mc}) and diffuse-ISM-type (Equation \ref{eq_n2av_diffuse})  medium, respectively.
The three $M_\mathrm{H_{2}}$ derived using radio and optical measurements agree well with each other.
The H$_{2}$ mass derived from CO is used for the discussion in the rest of the paper.

Although only upper limits could be achieved for many LSB objects,  they   largely follow the trend established by  star forming  galaxies towards the lower end in both $M_\mathrm{H_{2}}$ and SFR axes, consistent with the finding of \cite{Mcg17}.
On the other hand, quiescent galaxies, mostly early types,  have  lower SFR for a given  $M_\mathrm{H_{2}}$  and a lower CO detection rate than that of star-forming galaxies \cite[see also][]{Cal18}.
Due to the low SFR, Totoro deviates from the SFR-$M_\mathrm{H_{2}}$ relation formed by LSBs and star-forming  galaxies and appears to overlap with quiescent galaxies.

The H$\alpha$ emission in quiescent galaxies is dominated by  LI(N)ER excitation \citep{Hsi17,Pan18}.
In the LI(N)ER regions of quiescent galaxies, surface density of H$\alpha$ luminosity ($\Sigma_\mathrm{H\alpha}$) is found to be tightly correlated with underlying $\Sigma_\mathrm{\ast}$  \citep{Hsi17}, in which the H$\alpha$ are primarily powered by the hot, evolved stars rather than recent star formation.
BPT diagnostics suggest that Totoro is powered by a composite (LI(N)ER-HII mix) mechanism (\citetalias{Lin17}).
 If Totoro is  analogous to a  LI(N)ER region in quiescent (early-type) galaxies,  the average $\Sigma_\mathrm{H\alpha}$ of Totoro corresponds to a $\Sigma_\mathrm{\ast}$ of $\sim$ 7 $\times$ 10$^{8}$ M$_{\sun}$ kpc$^{-2}$  according to the  ``resolved LI(N)ER sequence''  of quiescent galaxies reported by  \cite{Hsi17}. 
 The predicted $\Sigma_{\ast}$ is higher than the mass of Satsuki's stellar halo by a factor of 3 -- 5; however, there is no distinct stellar counterpart at the location of Totoro.  
For this reason, in spite of the overlap in Figure \ref{fig_CO_ks}, Totoro is  not analogous to the  LI(N)ER region in quiescent galaxies.
 Lastly, we should note  that  if the true global SFR of Totoro is  considerably lower than the  upper limit, Totoro would  fall below the main cloud of data points of quiescent galaxies.
 
As a whole,  Totoro is unlikely to be consistent with   nearby normal star-forming, early-type, and low-surface-brightness galaxies in terms of the Kennicutt-Schmidt  relation and the resolved LI(N)ER sequence. In addition, it is worth noting that the  average $A_\mathrm{V}$ of Totoro corresponds to a SFR surface density ($\Sigma_\mathrm{SFR}$) at least 6 times higher than the observed upper limit  based on the  local $A_\mathrm{V}$-$\Sigma_\mathrm{SFR}$  relation derived from  MaNGA  galaxies \citep{Li19}. Generally speaking, the dust and star formation relation (if any) of Totoro is  also dissimilar to that of star-forming regions in nearby galaxies  even when the systematic dependencies on other physical properties (e.g., metallicity) are considered.

While LSBs have been studied for decades, recently, \cite{Van15} have identified a new class of LSBs  in the Coma cluster, ultra-diffuse galaxies (UDGs).
These UDGs have a surface brightness as low as $>$ 24.5 mag arcsec$^{-2}$, but their sizes are similar to those of $L^{\ast}$ galaxies.
It is not clear how UDGs were formed. 
One possible scenario is that UDGs are failed massive galaxies, which lost their gas  at high redshift by ram-pressure stripping or other effects after forming their first generation of stars \citep{Van15,Yoz15}.
Our finding of a large molecular gas reservoir appears  in direct conflict with this scenario.

On the other hand, several studies suggest that UDGs are extended dwarf galaxies. 
Some simulations predict them to be rapidly rotating \citep{Amo16}.
 \cite{Dic17} suggest that the extended sizes of UDGs are the consequence of strong gas outflows driven by starbursts.
 However,  the SFR of Totoro is extremely low and the  system is not rotating.
Altogether, there is no evidence  in our data to support Totoro as an UDG.

In summary, the newly obtained CO, H$\alpha$, and $u$-band data allow us to better constrain the nature of Totoro; however, combining the results in Section \ref{sec_sep_gal}, we  argue that Totoro is unlikely to be a separate galaxy interacting with the dry merger  (Satsuki and Mei).
The reasons include the lack of  stellar counterpart and tidal features and the  different star formation, ionized and cold gas, and dust properties (i.e.,  $M_\mathrm{H_{2}}$-SFR, $\Sigma_{\ast}$-$\Sigma_\mathrm{H\alpha}$, and $A_\mathrm{V}$-$\Sigma_\mathrm{SFR}$ relations and gas kinematics) from that of a variety of nearby galaxy populations (i.e., star-forming and quiescent galaxies, LSBs, and UDGs).

\subsection{AGN-driven Activity} 
\label{sec_agn}

\subsubsection{Multi-wavelength Nuclear Characteristics}

{\bf \emph{Radio}}. A possible origin of Totoro is gas that is photoionized by X-ray emission from a misaligned blazar at the core of Satsuki.
A blazar is a  sub-class   of  radio-loud AGN which   is characterized by one-sided    jet structure \citep{Urr95}.
However, blazars  invariably have  bright compact radio cores,  whereas no detection at 1.4 GHz is found for Satsuki \citep{Wan19,Osu19} and there is only marginal detection at 5 GHz (\citetalias{Lin17}).
For these reasons, a blazar is a very unlikely scenario for Totoro.

 {\bf \emph{Optical}}. The H$\alpha$ equivalent width (EW)  is $<$  3\AA\, across the entire H$\alpha$-emitting region and the EW does not present a rise towards the center of Satsuki. 
 The H$\alpha$ velocity dispersion does not present a central peak either (\citetalias{Lin17}).
 Moreover, the H$\alpha$ luminosity of Satsuki follows the continuum emission, i.e., does not present a central peak emission with stronger intensities that follows an $r^{-2}$ or lower decline from the center \citep[][]{Sin13}. 
Therefore, Satsuki presents a lack of optical characteristics of a strong AGN.

\begin{figure*}
	\centering
	\includegraphics[width=0.9\textwidth]{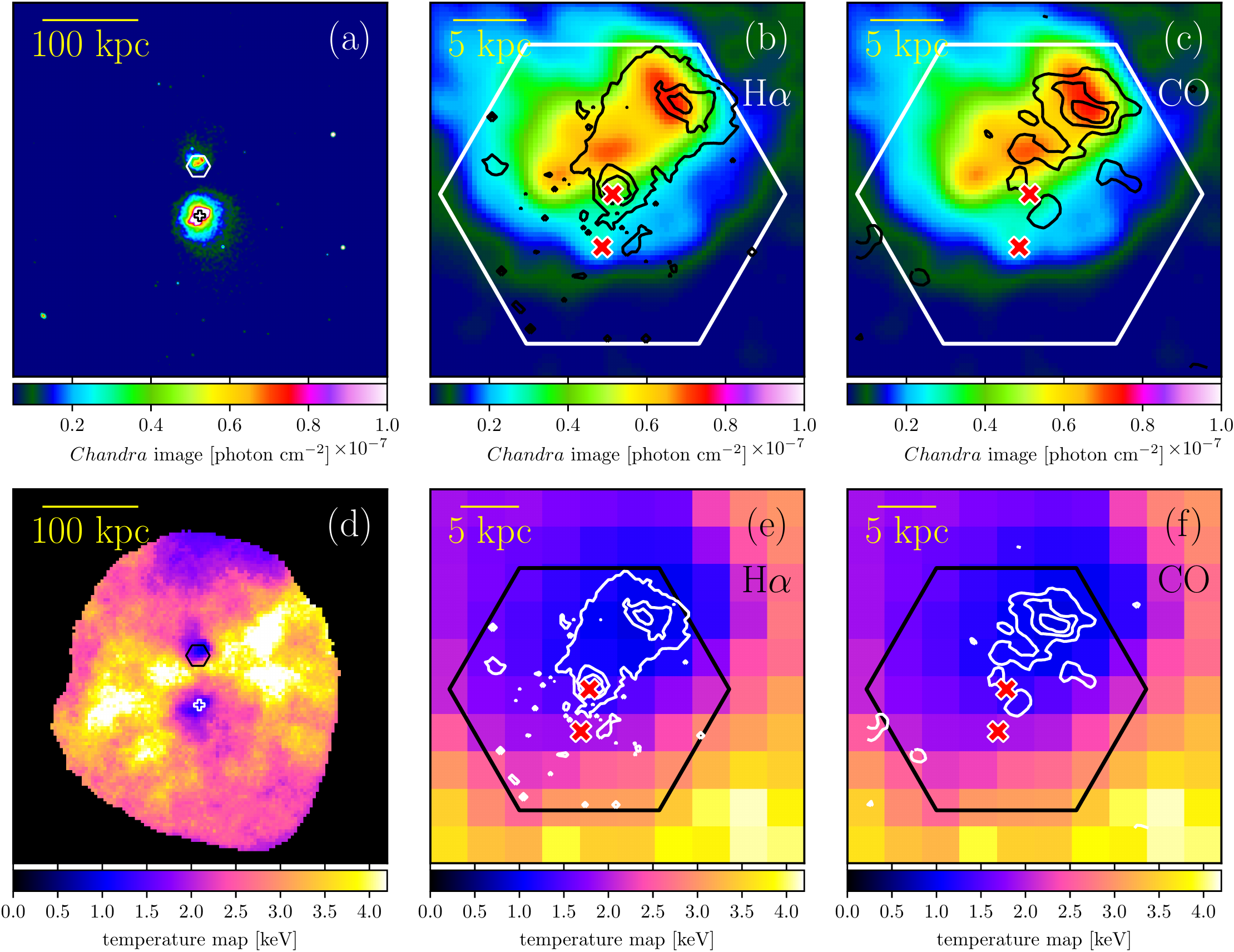}  
	\caption{\emph{Chandra}  0.5 -- 2 keV maps (a)-(c) and temperature maps (d )-(f) taken from \cite{Osu19}. In each row, from left to right we show the full map, zoom in on the MaNGA FoV with H$\alpha$ contours overlaid, and zoom in on the MaNGA FoV with CO contours overlaid. In panel (a) and (d), the nucleus of NGC 6338 is marked  by an open cross. The  MaNGA hexagonal  FoV is also indicated in the figures. In panel (b), (c), (e), and (f), the  red filled crosses indicate the nuclei of Satsuki and Mei.  The maps show that the  H$\alpha$ and CO emissions are coincident with X-ray structures and  low temperature regions.}
	\label{fig_Chandra}
\end{figure*}

{\bf \emph{X-ray}}. A comprehensive study of the X-ray emission of the region based on \emph{Chandra} and \emph{XMM-Newton} data has been reported recently by \cite{Osu19}.
Figure \ref{fig_Chandra}a  shows the \textit{Chandra} X-ray image of the NGC 6338 group taken from \cite{Osu19}. 
X-ray and galaxy velocity studies have shown the group to be a high-velocity near head-on merger \citep{Wan19}, with the two group cores visible as bright clumps of X-ray emission with trailing X-ray tails. The northern and southern clumps are respectively associated with VII Zw 700 and NGC~6338.
Figure \ref{fig_Chandra}b zooms in to the MaNGA observed region.
The northern X-ray clump, around Satsuki, is dominated by a bar of X-ray emission extending roughly southeast-northwest. The clump and bar are centered somewhat to the north of the optical centroid of Satsuki, and the bar is made up of three X-ray knots whose intensities progressively decrease from the western end. The nuclei of Satsuki and Mei are not correlated with the brightest X-ray emission \citep[see also][]{Wan19}.
Nonetheless, after subtracting the overall surface brightness distribution, there is a hint of excess X-ray emission at the position of the nuclei of Satsuki and Mei (see \citealt{Osu19} for the construction of the overall surface brightness distribution).
X-ray luminosities of 4.17 $\times$ 10$^{39}$ erg s$^{-1}$ for the nucleus of Satsuki   and $\leqslant$ 2.5 $\times$ 10$^{38}$ erg s$^{-1}$ for  Mei are reported \citep{Osu19}.
The values suggest that there are no strong X-ray AGN  in Satsuki or Mei or the AGNs are fading.
For reference, the X-ray luminosity of Totoro is (4.4$\pm$0.2) $\times$ 10$^{40}$ erg s$^{-1}$.

\subsubsection{Gas  Ejected by AGN}
\label{sec_agn_out}
All in all, the multi-wavelength data show no direct evidence for an  active ongoing AGN in Satsuki.
However, it is known that the AGN luminosity can vary over timescales   as short as 10$^{5}$ years. 
Thus, we still cannot rule out the possibility of a recent  AGN outflow.
AGN-driven extended outflows are detected in multiple phases, including ionized gas and many molecular lines \citep[][and references therein]{Rob20}.
Even though the observed radial extent of AGN outflows is  $\leq$ 1 kpc in most  cases,    galactic scale outflows, as Totoro would be, given its extreme distance from Satsuki, are also reported \citep[e.g.,][]{Lop19,Leu19,Lop20}.
 However, the low gas velocities (Figure \ref{fig_CO_flux_vel}c and  \ref{fig_CO_flux_vel}d) do not support the scenario of an energetic outflow.

Nonetheless, we should note that although an AGN does not seem like a plausible mechanism for moving H$\alpha$ or CO out of the galaxy core to the current position of Totoro, the potential cavities identified in Satsuki suggest that in the past ($\sim$ 40 Myr ago) the AGN was indeed active and had non-negligible impact on its surroundings \citep{Osu19}.

\subsubsection{Gas  Ionized by AGN}

There have been studies showing that AGNs are able to
ionize gas  extending to large distance, such as the well-known ionized cloud Hanny's Voorwerp $\sim$ 20 kpc from its host galaxy IC~2497  \citep[e.g.,][]{Hus08,Lin09,Hus10}.
It is believed that an interaction-triggered,  currently fading  AGN, illuminated and ionized Hanny's Voorewerp \citep{Lin09,Joz09,Kee12}.
Diffuse ionized gas similar to Hanny's Voorewerp was also found $\sim$ 32 kpc north of the   iconic interacting galaxies NGC 5194/5195 or M51 \citep{Wat18}.
The  low AGN  luminosity and  activity and the  tidal history provide  similarities between Satsuki, IC~2497 and M51.


However, there are also several differences between Hanny's Voorwerp  and the M51 cloud  and Totoro in terms of excitation state (Seyfert versus LINER-HII, morphology (diffuse for Hanny's Voorwerp and M51's cloud versus centrally-concentrated for Totoro), and the properties of the host galaxy  (late type versus early type, see \citetalias{Lin17} for the details).
For these reasons, we argue that this scenario is unlikely.

\subsection{Cooling Gas}
\label{sec_cooling}
\subsubsection{Spatial Comparison of Cold, Warm, and Hot Gas}
\label{sec_cooling_spatial}
A scenario we did not consider in \citetalias{Lin17} is the cooling of the hot IGM or ICM. Observations of the central regions of some galaxy groups and clusters show strong X-ray emission suggesting that the IGM and ICM are undergoing rapid cooling \citep[e.g.,][]{Fab94}, and in some cases ionized and molecular gas which are thought to be the product of that cooling  \citep{Bab18,Lak18,Oli19,Rus19}.

 \cite{Osu19} shows that the X-ray peak of the southern clump in Figure \ref{fig_Chandra}a is consistent with the optical centroid of NGC~6338. Three X-ray filaments are observed to extend from the galaxy center,  following the same branching   filamentary structure as the H$\alpha$ gas (Figure \ref{fig_Ha_large}c).
These X-ray filaments are cooler than their surroundings and have very low gas entropies and short cooling times \citep[][]{Osu19}, strongly indicating that they are a locus of cooling from the IGM.
Young X-ray cavities are also found in NGC 6338, suggesting recent AGN outbursts in this galaxy. 

As shown in Figure \ref{fig_Chandra}b, the peak of H$\alpha$ emission (contours) at the nucleus of Satsuki is not associated with the bar, but the position of Totoro corresponds to the brightest X-ray peak. 
This was first noted by \cite{Osu19} who showed that this was also the coolest part of the X-ray bar, and concluded that the H$\alpha$ was likely material cooled from the IGM, as in the filaments of NGC~6338. 
They also suggested that the offset between Satsuki and the center of the X-ray clump is evidence that ram-pressure forces caused by the (supersonic) motion of the galaxy are detaching the gas from the galaxy. 
In this scenario, the X-ray bar would once have been centered on Satsuki, but has been pushed back to the north and perhaps along the line of sight, away from the galaxy core. 


 Figure \ref{fig_Chandra}c shows our CO data overlaid on the \textit{Chandra} image. The molecular gas is well correlated with the western X-ray knot, and (as previously discussed) with the H$\alpha$. Figures \ref{fig_Chandra}d-e  show the \textit{Chandra} temperature maps from \cite{Osu19} showing the cool gas around NGC~6338 and Satsuki, and the high-temperature gas between the two, which has been shock-heated by the group-group merger. Overlaid H$\alpha$ and CO contours show  that the ionized and molecular gas are located in the coolest part of the IGM, as in NGC~6338 and the centers of other groups and clusters  \citep[e.g.,][]{Sal06,Ham12}.

\subsubsection{Cooling Time and Gas Properties}
\label{sec_cooling_time}
Studies have shown that warm and cold gas in the brightest group and cluster galaxies (BCGs)  are preferentially  observed   in systems where the  cooling times lie below $\sim$ 1 Gyr \citep{Edg01,Sal03,Cav08,Raf08,Pul18}.
\cite{Osu19} found that the cooling times are as short as  $<$1~Gyr in both cores of Satsuki and NGC~6338.
We can also estimate the cooling time around Totoro, i.e., the west end of the X-ray bar as a cylinder of radius 2.8$\arcsec$(1.5 kpc) and length 11.2$\arcsec$ (5.9 kpc, corresponding to the Region 2 in Figure 10b of \citealt{Osu19}).
The temperature, density, and luminosity of the X-ray gas are estimated via spectral fitting of the $Chandra$ data.
Using  Equation (1) in \cite{Osu19}, we estimate the cooling time ($t_\mathrm{cool}$) in the Totoro region to be  2.2$^{+0.2}_{-0.1}$ $\times$ 10$^{8}$ yr, well within  the regime where warm and cold gas are   expected.

If the warm and cold  gas have indeed cooled from the hot gas, one would expect the hot gas to be  significantly more massive than the cool/warm gas.
The hot gas mass is derived from a radial deprojected profile (Figure 9 in \citealt{Osu19}), centering on the  middle of the X-ray bar.
Within a radius of 6.5 kpc,  the hot gas mass ($M_\mathrm{X}$) is $1.2_{-0.24}^{+1.07}$ $\times$ 10$^{9}$ M$_{\sun}$.
The derived hot gas mass is indeed significantly higher than the total mass of warm ($\sim$ 10$^{5}$ M$_{\odot}$ from H$\alpha$) and cold gas  ($\sim$ 10$^{8}$ M$_{\odot}$ from CO).

The ratio of the cold to hot gas of Totoro is $\sim$ 17\%.
\cite{Pul18}  investigate the molecular gas properties of 55 central cluster galaxies. 
They found a strong correlation of hot and cold gas mass traced by X-ray and CO, suggesting that the hot and cold gas arise from the same ensemble of clouds.
In other words, the cold gas is unlikely a result of external effects, such as merger or stripping from a plunging galaxy.
The average fraction of cold to hot gas in their sample is $\sim$ 18\%.
The cold to hot gas mass ratio of Totoro is in good agreement with that of  these central cluster galaxies \citep[see also][]{Oli19}.
The consistency  provides  support that a similar process to that in the central cluster galaxies has occurred in Satsuki and Totoro.

\begin{figure}
	\centering
	\includegraphics[width=0.43\textwidth]{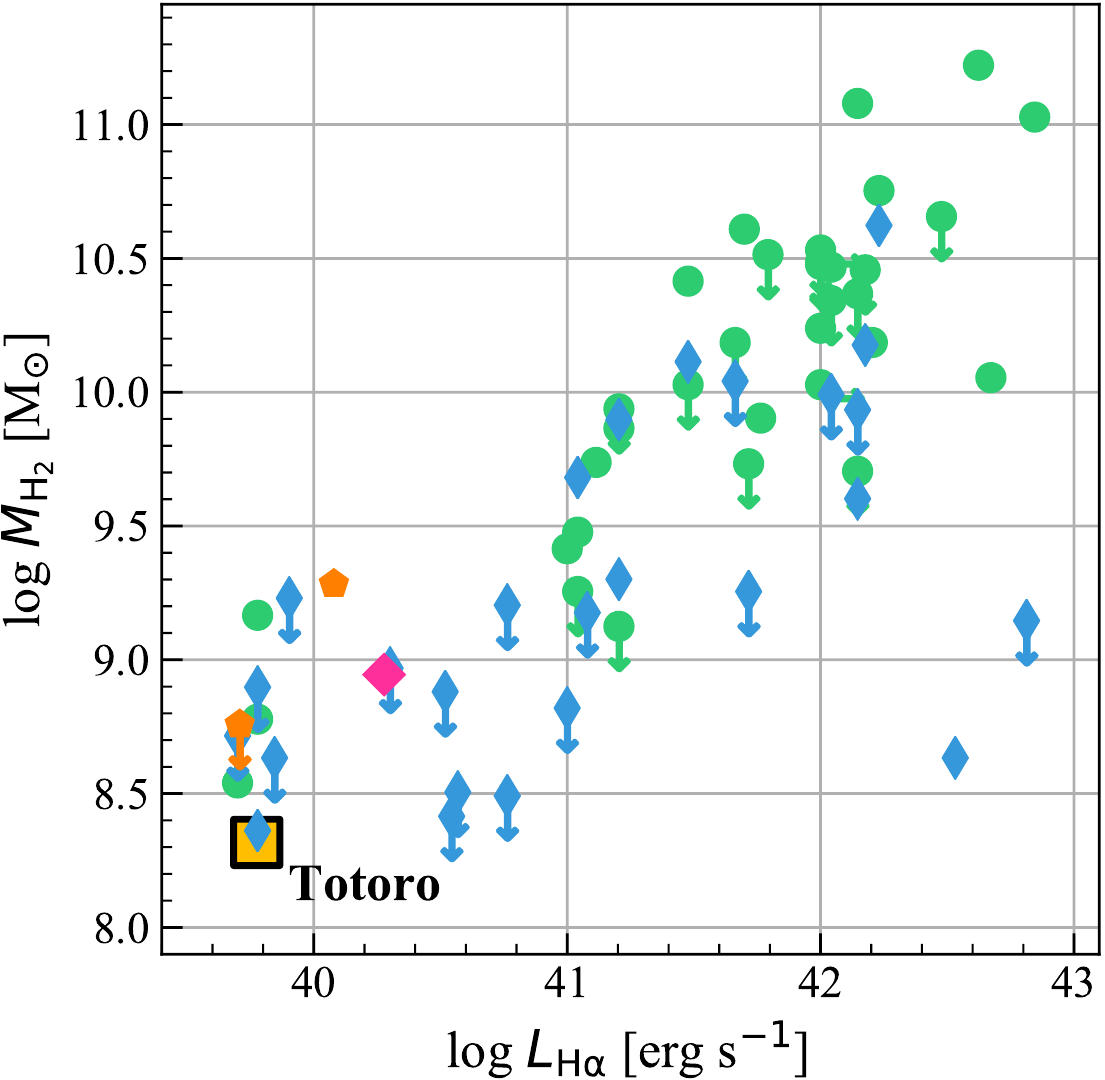}  
	\caption{H$\alpha$ luminosity versus molecular gas mass of cooling gas in cluster galaxies. Data points marked with  circles,  thin diamonds, pentagons, and diamond are taken from \cite{Edg01}, \cite{Sal03}, \cite{Mcd12}, and   \cite{Ham12},  respectively. Totoro is shown by an orange square and lies on the relationship of other systems.
	}
	\label{fig_Cooling}
\end{figure}

In addition, molecular  mass has been found to be correlated with the amount of H$\alpha$ gas expressed by $L_\mathrm{H\alpha}$ in the cooling-core galaxies of clusters  \citep{Edg01,Sal03}.
In Figure \ref{fig_Cooling} we show the H$_{2}$ gas mass versus $L_\mathrm{H\alpha}$ for data taken from the literatures. 
Totoro is overlaid with a orange square, and falls at the low  end of the correlation.
The consistency of Totoro  with the $M_\mathrm{H_{2}}$- $L_\mathrm{H\alpha}$ relationship of cooling systems again supports  a similar process of cooling in the system. 
To put it another way, our multi-wavelength data of Totoro show that its position on the mass (or luminosity) relations between cold ($\lesssim$ 100 K; CO), warm ($\sim$ 10$^{4}$ K, H$\alpha$), and hot ($>$ 10$^{7}$ K, X-ray) gas is in line with the gas content of systems with short cooling times. 
We summarize the physical properties of Totoro in Table \ref{Tab_totoro}.

As  a side note, the cold and warm gas  in cooling systems often appear  filamentary, but it is worth noting that the multi-phase gas morphologies of our target are strikingly similar to the BCG of cluster Abell 1991 reported by \cite{Ham12} (see their Figure 2).
The H$\alpha$-emitting gas is relatively circular (blob-like) and is  spatially coincident with  the most rapidly cooling region (X-ray peak) of the ICM.
The  peak in the H$\alpha$ and X-ray gas lies roughly 11 kpc to the north of the BCG, and there are connecting arm structures between the peak and the secondary peak at the galactic nucleus.
Moreover,   the bulk of the molecular gas with a mass of $\sim$ 8$\times$ 10$^{8}$ M$_{\sun}$ is also found at the location of the  cooling region. 
However, note that spatial resolution must play an important role in detecting the  morphology of cooling gas.
We cannot rule out that the morphologies of Totoro and the cooling gas in  Abell 1991 are  more filamentary than a single peak.

Finally, the velocity fields of cooling gas in galaxy clusters, traced by CO and H$\alpha$,  are characteristic of  slow  motions (projected velocity $<$ 400 km s$^{-1}$), narrow line widths ($<$ 250 km s$^{-1}$)  and  a  lack  of  relaxed (e.g., rotating) structures \citep[e.g.,][]{Sal08,Ham12,Oli19}.
The observed velocity structures, despite low velocity resolutions, of  Totoro traced by CO and H$\alpha$ (Figures \ref{fig_CO_flux_vel}c and \ref{fig_CO_flux_vel}d) agree with that of other cooling systems. 
In addition, the gas velocities of Totoro traced by  CO and H$\alpha$ are not exactly identical, but the differences are small, $\sim$ 35 km s$^{-1}$ at the main blob region and 60 -- 90 km s$^{-1}$ at the connecting arms.
This is also consistent with the finding by \cite{Oli19} that  the velocity difference between CO and H$\alpha$ gas  is  well  below 100 km s$^{-1}$,  providing an additional support for CO and  H$\alpha$ gas arising from the same bulk of clouds.
The velocity difference may be related to different velocity resolution of CO and H$\alpha$ observations and line of sight projection effect.
We should note again that our CO and H$\alpha$ observations suffer from low velocity resolutions, therefore the velocity fields must be interpreted  with caution. Nonetheless, the maps  still provide a guide of the velocity resolution needed for future observation of this system.
Future high velocity-resolution CO and H$\alpha$  observations are  required to reveal the detailed gas kinematics of Totoro.


\subsubsection{Environments}
\label{sec_cooling_env}
While most of the studies of cooling gas focus on BCGs in rich clusters,  cold gas cooling from the hot X-ray medium is  also observed at smaller scales in galaxy groups, such as  NGC 5044, NGC  4638  and  NGC5846 \citep{Dav14,Tem18}.
In fact, the observed cool-core fractions for galaxy groups are slightly higher than those of  galaxy clusters \citep{Osu17}.
Therefore,  gas cooled from the IGM is not unique  to Totoro.

In Section \ref{sec_ram}, we argue that the cold gas in Totoro is unlikely to be the primitive gas of Satsuki being stripped by ram-pressure. 
Nonetheless, the X-ray tail (Figure \ref{fig_Chandra}a) and the offset between the center of the X-ray bar and the optical centroid of Satsuki are both evidence that the motion of the dry merger  (Satsuki and Mei) is rapid enough to lead to stripping of the hot gas halo \citep{Osu19}. 
The question then arises of where and when the ionized and molecular gas we observe was formed; have ram-pressure or other effects changed its location? 
We might normally expect to see the most rapid cooling in or near the galaxy center. 
However, the offset of the X-ray bar means that the densest, coolest IGM gas is no longer located at the center of Satsuki. 
Ram-pressure, by pushing the  X-ray halo and bar away from the core of Satsuki, may therefore have caused a reduction in cooling in the galaxy center \citep{Osu19}.

CO emission occurs in dense molecular clouds which are ``self-shielding'' from the ionizing effects of the surrounding environment. 
While some of the H$\alpha$ emission likely comes from the outer layers of such molecular cloud complexes, observations show that in some cooling systems the H$\alpha$ emission is considerably more extended than the molecular gas \citep[e.g., in NGC~5044; ][]{Sch20} which may suggest it is associated with a less dense cooled gas component.
 Such low-density material would likely move with the surrounding IGM if ram-pressure pushed it back from the galaxy. Dense molecular clouds would not be affected by ram-pressure, and might be expected to fall under gravity toward the center of Satsuki, unless they are connected to the IGM via magnetic fields \citep{Mcc15} or surrounding layers of neutral and ionized gas \citep{Li18}. Even with such connections to the surrounding environment, it seems implausible that the molecular gas could have condensed out of the IGM in the core of Satsuki and then been uplifted. The correlation between the CO, H$\alpha$ and the coolest X-ray gas strongly suggests that the molecular and ionized gas is the product of cooling from the IGM at its current location, i.e., that cooling has occurred (and may be ongoing) in Totoro, well outside the center of Satsuki.

 Last but not least, as suggested by \cite{Osu19} and this work,  we are witnessing  a merger between two groups undergoing rapid radiative cooling.
 Further analysis on the multi-wavelength phase of cooling gas in NGC 6338, from cold to hot gas as in this study,  will be carried in a separate paper (O'Sullivan et al. in preparation). 

%




\subsubsection{Future of the Gas}
\label{sec_cooling_future}
Cooling gas can potentially serve as the fuel for an AGN and/or central star formation \citep[][]{Ode08,Raf08,Mit09,Hic10,Fog17}.
In X-ray bright groups, like our target, star formation in the central galaxy is generally  weak even in systems known to be cooling. 
By contrast, as many as 85 -- 90\% of group-dominant galaxies have radio AGN, moreover, dominant galaxies with active or recently active radio jets are relatively common in X-ray bright groups \citep{Kol18,Kol19}.  
Galaxy cluster-dominant galaxies, however,  seem much more likely to have significant star formation.


Here we consider whether the cooling gas would fuel star formation  assuming the gas will fall back into Satsuki.
\cite{Mcd18} compare  the  cooling rate of the ICM/IGM to the observed SFR in the central galaxy for a  sample of isolated ellipticals, groups, and clusters.
They found that   the cooling ICM/IGM  is not providing the fuel for star formation in systems with cooling rate $<$ 30 M$_{\sun}$ yr$^{-1}$,  which are  dominated by groups and isolated ellipticals.
On the other hand,  SFR increases with increasing cooling rate for the rapidly cooling systems ($>$ 30 M$_{\sun}$ yr$^{-1}$), presumably  due to an increase in either  the cooling efficiency of the hot gas or the star formation efficiency of the cooled gas \citep[see also][]{Edg01,Sal03,Ode08}.
The cooling rate ($M_\mathrm{X}$/$t_\mathrm{cool}$) of Totoro is  $\sim$ $1.2$ $\times$ 10$^{9}$  M$_{\sun}$/2.2 $\times$ 10$^{8}$ yr $\approx$ 5  M$_{\sun}$ yr$^{-1}$.
In this aspect,  the cooling gas in our group-dominant, relatively low  cooling-rate system  is less likely to significantly contribute to star formation.

Moreover, star formation can be suppressed by AGN feedback even  in  systems with short cooling times.
The summed  AGN jet  power ($P_\mathrm{cav}$) for both cavities associated with Satsuki from \cite{Osu19} is  (0.97 -- 2.67) $\times$ 10$^{41}$  erg s$^{-1}$  (an estimate of heating; the actual value depends on the cavity age used: buoyant rise time, sonic expansion time-scale, or refill time), the energy is comparable to the  X-ray luminosity  for the whole X-ray emitting gas ($\sim$ 3 $\times$ 10$^{41}$ erg s$^{-1}$; an estimate of cooling).
\cite{Raf08} find a tendency for star-forming systems to have low $P_\mathrm{cav}$/$L_{X}$ ratios ($<$ 1)  and quiescent systems to have high $P_\mathrm{cav}$/$L_{X}$ ratios  ($>$ 1),  supporting the suppression of star formation  by AGN feedback.
However, this is not exclusively the case; star-forming cooling systems can have high $P_\mathrm{cav}$/$L_{X}$ ratios, and vice versa.
The estimated heating available from the cavities around Satsuki make it a borderline case, and the ongoing merger and stripping add to the complexities. The fate of Totoro may depend on how effectively energy from the cavities can heat their surroundings.

Finally, it is worth mentioning that Satsuki  and Totoro may host a little star-formation activity with an upper limit of $<$ 0.059 and $<$ 0.047 M$_{\sun}$ yr$^{-1}$, respectively, assuming all the H$\alpha$ fluxes come from star formation.
It is unclear  if the current star formation (if any) is related to the cooling and gas fueling processes.
\cite{Mcd18} attribute the  low-level star formation in low cooling rate systems to  recycling of gas lost by evolved stars, namely,  the star formation is not related to  cooling gas.

\section{Summary}
\label{sec_summary}
In  \citetalias{Lin17}, we  identified an H$\alpha$ blob Totoro $\sim$ 8 kpc away from a dry merger  (Satsuki and Mei) from MaNGA data (Figure \ref{fig_clusters}). 
Here we present new optical (wide-field H$\alpha$ and $u$-band), millimeter ($^{12}$CO(1-0)) observations, and published X-ray data \citep{Osu19}, with the aim of providing significant constraints and answers to fundamental questions regarding the nature of Totoro.
The main conclusions of this paper are as follows:
\begin{itemize}
\item The data disfavor the scenario that Totoro is stripped from  Satsuki by ram-pressure based on the morphology   and kinematics of ionized (H$\alpha$) and molecular gas) and the properties of the host galaxy (Section \ref{sec_ram} and Figure \ref{fig_CO_flux_vel}).
\end{itemize}

We consider whether Totoro is a separate galaxy interacting with the dry merger  (Satsuki and Mei) from several aspects:

\begin{itemize}
\item We apply three commonly-used methods to   $g$-, $r$-, and $i$-band images (Figure \ref{fig_CFHT_all}) to look for  an underlying stellar counterpart of Totoro. However, we find no compact underlying stellar component associated with Totoro (Section \ref{sec_gal_stellar} and Figure \ref{fig_lightfitting}). 

\item No tidal tail feature is seen in H$\alpha$ beyond the MaNGA FoV. If Totoro is a galaxy interacting with the dry merger  (Satsuki and Mei), it  may have a non-typical tidal history  and morphology, or it is a completely disrupted   low-surface-brightness dwarf galaxy (Section \ref{sec_gal_ion}, Figure  \ref{fig_Ha_large} and Table \ref{tab_cfht_mag}).



\item  However, Totoro shows different star formation, gas, and dust properties (in terms of  $M_\mathrm{H_{2}}$-SFR, $\Sigma_{\ast}$-$\Sigma_\mathrm{H\alpha}$, and $A_\mathrm{V}$-$\Sigma_\mathrm{SFR}$ relations and gas kinematics) from that  of a variety of nearby galaxy populations (i.e., star-forming and quiescent galaxies, low-surface-brightness and ultra-diffuse galaxies). Therefore, Totoro is unlikely to be a separate galaxy interacting with the dry merger (Satsuki and Mei) (Section \ref{gas_sfr_mh2} and  Figure \ref{fig_CO_ks}).


\item The $u$-band data, which are sensitive to recent star formation, show no strong sign of   recent star formation  at the position of Totoro. Therefore, the ionized gas of Totoro is unlikely to be powered by star formation, confirming the results of emission line ratios diagnostics in  \citetalias{Lin17}  and previous bullet-point that Totoro is not an analogue of a star-forming region in  nearby galaxy populations  (Section \ref{sec_gal_stellar} and Figure \ref{fig_lightfitting}).
\end{itemize}

We consider whether Totoro is a result of  AGN activity.

\begin{itemize}
\item   However,  in spite of possible past AGN outbursts, the multi-wavelength data show no direct evidence for an active ongoing AGN in Satsuki or Mei. Moreover,  Totoro is unlikely to be  gas being ionized or ejected by an AGN  as its physical properties (gas excitation state, morphology,  and kinematics, etc.) are distinct from similar objects  (Section \ref{sec_agn}).
\end{itemize}

Finally, we consider whether  Totoro is formed via cooling of hot IGM   as implied by \cite{Osu19}.
\begin{itemize}
\item We compare the spatial distribution of H$\alpha$ and CO with X-ray intensity and temperature maps.
We find that the ionized and molecular gas are related to the most rapidly cooling region of the hot IGM.
The cooling time in the Totoro region is  well within the regime where cooling is expected (Section \ref{sec_cooling_spatial} and  \ref{sec_cooling_time} and Figure \ref{fig_Chandra}). 
\end{itemize}

\begin{itemize}
\item The mass (or luminosity) relations between cold ($<$ 100 K; CO), warm ($\sim$ 10$^{4}$ K, H$\alpha$), and hot ($>$ 10$^{7}$ K, X-ray) gas,  as well as gas kinematics are in line with the gas content of cooling systems, supporting again that Totoro originates from the same physical process of cooling gas (Section \ref{sec_cooling_time} and Figure \ref{fig_Cooling}).  
\end{itemize}

\begin{itemize}
 \item   Previous study by \cite{Osu19} suggests that the densest, coolest X-ray gas has been pushed away from the core of the host galaxy Satsuki by ram-pressure. The correlation between the CO, H$\alpha$ and the coolest X-ray gas presented in this work strongly suggests that the molecular and ionized gas is the product of cooling from the hot X-ray gas  and is formed at its current location, leading to the observed \emph{offset} cooling and the reduction in cooling in the galaxy core (Section \ref{sec_cooling_env}). 
 
  \item  The cooling rate of Totoro is considerably lower than that of star-forming cooling systems. The estimated heating available from the
cavities around Satsuki is comparable to the cooling X-ray luminosity. The fate of Totoro may depend on how effectively energy from the cavities can heat their surroundings, but note that the ongoing merger (Satsuki, Mei, and NGC~6338) and stripping add to the complexities  (Section \ref{sec_cooling_future}).
\end{itemize}
In the  majority of clusters the peaks of optical and X-ray emission are very close to the center of galaxy, so it is difficult to determine whether  the appearances of warm or even  cold gas  are primarily related to the cooling of the IGM/ICM or the host galaxy.
Offset cooling is rare \citep[$<$ 5\%, e.g.,][]{Ham12},  therefore VII Zw 700 provides an exceptional opportunity to constrain the gas cooling process and the interplay between  cooling gas and  host galaxy.
In future work, we intend to constrain the gas kinematics  at the connecting arms with high-spectral-resolution data to quantify the potential of  a fueling process for Satsuki. A detailed multi-wavelength analysis, from CO to H$\alpha$, to other optical lines (e.g., H$\beta$, [NII], [OIII], etc.), and to X-ray, will be presented in an upcoming paper (O'Sullivan et al in preperation).
Moreover, the large sample of optical IFU survey MaNGA  ($\sim$ 5,000 galaxies   in the latest SDSS data releases DR15/16 and  $\sim$ 10,000 in the future)  along with the \emph{Chandra} archive   is a  suitable starting point  for future  multi-wavelength statistical studies of cooling gas properties.
Although MaNGA does not target specific environments, the large sample size of MaNGA ensures observations of numerous galaxies located in groups and clusters. 
Given the significant fraction of cooling cores in galaxy clusters/groups \cite[e.g.,][]{Mit09,Osu17}, more cooling gas candidates are expected in the final MaNGA sample.

\acknowledgments

We would like to thank the anonymous referee for constructive comments that  helped to improve the manuscript.
H.A.P thanks Eva Schinnerer, Christine Wilson, and Toshiki Saito for useful discussions. 
This work is supported by the Academia Sinica under the Career Development Award CDA-107-M03 and the Ministry of Science \& Technology of Taiwan under the grant MOST 108-2628-M-001-001-MY3.
M.J.M.~acknowledges the support of the National Science Centre, Poland
through the SONATA BIS grant 2018/30/E/ST9/00208, the Royal Society of
Edinburgh International Exchange Programme, and the hospitality of the
Academia Sinica Institute of Astronomy and Astrophysics (ASIAA).
EOS gratefully acknowledges the support for this work provided by the National Aeronautics and Space Administration(NASA) through \emph{Chandra} Award Number G07-18162X, issued by the  \emph{Chandra} X-ray Center, which is operated by the Smithsonian Astrophysical Observatory for and on behalf of NASA under contract NAS8-03060.
SFS is grateful for the support of a CONACYT grant  FC-2016-01-1916, and funding from the PAPIIT-DGAPA-IN100519 (UNAM) project.
J.G.F-T is supported by FONDECYT No. 3180210 and Becas Iberoam\'erica Investigador 2019, Banco Santander Chile.

This project makes use of the MaNGA-Pipe3D dataproducts \citep{San16b,San18}. We thank the IA-UNAM MaNGA team for creating this catalogue, and the Conacyt Project CB-285080 for supporting them.

This work is partly based on observations carried out under project number S16BE001 with the IRAM NOEMA Interferometer. IRAM is supported by INSU/CNRS (France), MPG (Germany) and IGN (Spain).

This work is partly based on observations obtained with MegaPrime/MegaCam, a joint project of CFHT and CEA/DAPNIA, at the Canada-France-Hawaii Telescope (CFHT) which is operated by the National Research Council (NRC) of Canada, the Institut National des Sciences de l'Univers of the Centre National de la Recherche Scientifique of France, and the University of Hawaii. 
The authors also wish to recognize and acknowledge the very
significant cultural role and reverence that the summit of Maunakea has always had within the indigenous Hawaiian community. We are most fortunate to have the opportunity to conduct observations from this mountain.

This work is partly based on observations obtained with  the 6-m telescope of the Special Astrophysical Observatory of the Russian Academy of Sciences  carried out with the financial support of the Ministry of Science and Higher Education of the Russian Federation (including agreement No. 05.619.21.0016, project ID RFMEFI61919X0016).
The analysis of the ionized gas distribution according SCORPIO-2 data  was supported by the grant of Russian Science Foundation project 17-12-01335 ``Ionized gas in galaxy discs and beyond the optical radius''.

This work use GAIA to derive aperture photometry.
GAIA is a derivative of the SKYCAT catalogue and image display tool, developed as part of the VLT project at ESO. SKYCAT and GAIA are free software under the terms of the GNU copyright. The 3D facilities in GAIA use the VTK library.

This research made use of APLpy, an open-source plotting package for Python \citep{Rob12}.

Funding for the Sloan Digital Sky Survey IV has been provided by the Alfred P. Sloan Foundation, the U.S. Department of Energy Office of Science, and the Participating Institutions. SDSS-IV acknowledges
support and resources from the Center for High-Performance Computing at
the University of Utah. The SDSS web site is www.sdss.org.

SDSS-IV is managed by the Astrophysical Research Consortium for the 
Participating Institutions of the SDSS Collaboration including the 
Brazilian Participation Group, the Carnegie Institution for Science, 
Carnegie Mellon University, the Chilean Participation Group, the French Participation Group, Harvard-Smithsonian Center for Astrophysics, 
Instituto de Astrof\'isica de Canarias, The Johns Hopkins University, Kavli Institute for the Physics and Mathematics of the Universe (IPMU) / 
University of Tokyo, the Korean Participation Group, Lawrence Berkeley National Laboratory, 
Leibniz Institut f\"ur Astrophysik Potsdam (AIP),  
Max-Planck-Institut f\"ur Astronomie (MPIA Heidelberg), 
Max-Planck-Institut f\"ur Astrophysik (MPA Garching), 
Max-Planck-Institut f\"ur Extraterrestrische Physik (MPE), 
National Astronomical Observatories of China, New Mexico State University, 
New York University, University of Notre Dame, 
Observat\'ario Nacional / MCTI, The Ohio State University, 
Pennsylvania State University, Shanghai Astronomical Observatory, 
United Kingdom Participation Group,
Universidad Nacional Aut\'onoma de M\'exico, University of Arizona, 
University of Colorado Boulder, University of Oxford, University of Portsmouth, 
University of Utah, University of Virginia, University of Washington, University of Wisconsin, 
Vanderbilt University, and Yale University.

\end{CJK}
\end{document}